\documentclass[review]{elsarticle}
\usepackage{url}
\usepackage{amsmath,amssymb}
\usepackage{algorithm}
\usepackage[noend]{algpseudocode}
\usepackage{graphicx}
\usepackage{subcaption}
\usepackage[margin=1in]{geometry}
\usepackage{multirow,tabularx}
\usepackage{rotating}
\usepackage{float}
\usepackage{chngcntr}

\newcolumntype{Y}{>{\centering\arraybackslash}X}
\renewcommand{\arraystretch}{0.9}

\newcolumntype{L}[1]{>{\raggedright\let\newline\\\arraybackslash\hspace{0pt}}m{#1}}
\newcolumntype{C}[1]{>{\centering\let\newline\\\arraybackslash\hspace{0pt}}m{#1}}
\newcolumntype{R}[1]{>{\raggedleft\let\newline\\\arraybackslash\hspace{0pt}}m{#1}}

\newcommand{\argmin}{\operatornamewithlimits{arg\ min}}
\makeatletter
\renewcommand{\Function}[2]{%
	\csname ALG@cmd@\ALG@L @Function\endcsname{#1}{#2}%
	\def\jayden@currentfunction{#1}%
}
\newcommand{\funclabel}[1]{%
	\@bsphack
	\protected@write\@auxout{}{%
		\string\newlabel{#1}{{\jayden@currentfunction}{\thepage}}%
	}%
	\@esphack
}
\makeatother

\begin{document}
\begin{frontmatter}
\title{Analyzing the performance of distributed conflict resolution among autonomous vehicles}
\author{\'Italo Romani de Oliveira}
\address{Department of Computer Engineering and Digital Systems, School of Engineering, University of S\~ao Paulo. Av. Prof. Luciano Gualberto, trav. 3, 158, CEP 05508-010}
\begin{abstract}
	This paper presents a study on how cooperation versus non-cooperation, and centralization versus distribution impact the performance of a traffic game of autonomous vehicles. A model using a particle-based, Lagrange representation, is developed, instead of a Eulerian, flow-based one, usual in routing problems of the game-theoretical approach. This choice allows representation of phenomena such as fuel exhaustion, vehicle collision, and wave propagation. The elements necessary to represent interactions in a multi-agent transportation system are defined, including a distributed, priority-based resource allocation protocol, where resources are nodes and links in a spatial network and individual routing strategies are performed. A fuel consumption dynamics is developed in order to account for energy cost and vehicles having limited range. The analysis shows that only the scenarios with cooperative resource allocation can achieve optimal values of either collective cost or equity coefficient, corresponding respectively to the centralized and to the distributed cases. 
\end{abstract}
\end{frontmatter}

\section{Introduction}
A transportation system operated by different entities can be understood as a multi-agent system with a variety of concerns and goals. There are different degrees of centralization in the organization of these systems, varying from highly centralized and hierarchical ones such as railroads, to very anarchical and decentralized ones such as country road traffic. The concept of ``anarchical'' is understood here as each agent maximizing its utility without sharing decision making with other agents and with little or no consideration of systemic information, besides being selfish. Air traffic can be considered a middle case because, on the one hand, it is strongly managed by Air Traffic Control Centers; but, on the other hand, it has to accommodate a lot of non-determinism in the flight  times, due to environmental and operational factors, and allows certain degrees of freedom for the pilot to choose its trajectory, as long as under proper coordination. Systems with loose coupling (i.e., with variable delay and hit rate caused by the human in the loop) limit the effectiveness that central coordination can achieve, making distributed control and self-organization a popular topic since many years \cite{Dimarogonas2003, Bakule2008}. Currently, with the expected increase in vehicle autonomy in various transportation modes, this becomes even more important. In the case of air traffic, approaches to decentralized or distributed control vary from the most radical ones \cite{RTCAFreeFlight, Hoekstra2001}, passing through some moderate concepts \cite{VilaplanaThesis,Moniz2009,Blom2015} and including more conservative ones, such as market-based mechanisms \cite{Waslander2008,Castelli,Schummer2013,Cruciol2015} which take place at the planning stages of the operations. The so-called Collaborative Decision Making \cite{Zellweger} in Air Traffic Management uses principles of negotiations but, as it is today, still needs hierarchy and a central authority. 

The benefits of distributed control are more flexibility to the pursuit of individual goals and more reliability in conflict solving, which emanates from the shorter communication paths and avoiding processing overload to the central entity \cite{Blom2014}, even though, it is less open to optimization than centralized solutions. Game theoretical studies show that huge performance improvements can be achieved in network traffic systems when coordinated solutions are implemented instead of anarchical ones \cite{Roughgarden02boundingthe,correa2005inefficiency}, but most of these studies take a Eulerian approach, whereby the successive vehicles' positions are not taken into account and the traffic flows steadily during a given period of time. The abstraction of vehicle positions makes it difficult modeling the temporal propagation of the traffic, localized traffic surges and phenomena such as vehicle collision and fuel exhaustion. These aspects need some way to model the vehicle displacement in the network, and a seminal reference on such type of modeling is~\cite{Daganzo1994}, which introduced the so-called Cell Transmission Model (CTM). In this model, a traffic way is divided into cells for which the counts of vehicles entering and leaving per unit of time are taken into account. Late evolutions of this model have been applied in combination with game theory~\cite{schadschneider2010stochastic, Tanimoto2015}, where the traffic is modeled by means of Cellular Automata (CA); however these CA case studies concern linear traffic and lane changes, not routing problems. Besides, the dominant approach of these works is related to statistical physics, which has a very different concept of equilibrium than that of game theory, despite these works making analogies with such theory. 

A remarkable study on congestion prediction in air traffic~\cite{Bayen2005} uses a fully Lagrangian model, where each vehicle is represented as a separate object, a hybrid automaton, for which a univocal trajectory is maintained. The present study is affine with such type of modeling but does not aim at predicting real traffic and, accordingly, uses a much simpler vehicle and traffic model. Another interesting aspect of~\cite{Bayen2005} is that it demonstrates the relations of its Lagrangian model with previous Eulerian models. And, because both these types of model should represent the same phenomena, it is possible as well to mix their features as it was done in~\cite{Sun2008}, where a Eulerian-Lagrangian model, named Large Capacity Cell Transmission Model or CTM(L), becomes the basis for applying mixed integer linear programming to minimize the total travel time, in a fully centralized manner. The results presented in this work show the great potential of CTM(L) for traffic optimization, however, these models need some complementation for dealing with non-determinism and prioritization of individual vehicles. 

The present work is devoted to efficiency and equity aspects of distributed control, the relevance of these topics coming from the following facts: first, one can observe the increasing importance of autonomous vehicles in several transportation modes, notably in ground transportation, with autonomous cars being developed by several vendors, and in aviation~\cite{Balin2016}, with the ever growing importance of drones; and, second, the concept of autonomous vehicle is usually associated with distributed control. As written in the same reference~\cite{Balin2016}, ``\emph{Autonomy is the ability to achieve goals while operating independently from external control}.'' Distributed control, however, can occur cooperatively or non-cooperatively, and this impacts the systemic traffic efficiency. When multiple autonomous vehicles use a protocol to resolve conflicts, there is some degree of cooperation, which can be as low as notifying the intent, and can be as high as agreeing to a conflict resolution algorithm and electing a leader to execute the algorithm and to direct the conflict resolution process. In the latter case, this cooperation becomes similar to centralized control, however, the uncertainties which this scenario retains on team formation and who will be elected as the leader still can be understood as a form of distributiveness. In this paper, however, the terms ``non-cooperation" and ``cooperation" refer to  cases of cooperation in different degrees of cooperativeness among vehicles, as it will be explained in section~\ref{sec:price_of_anarchy}.  

Recognizing the merits of the previous works on traffic modeling and optimization, cited above or else, the present paper aims at filling the existing void in applying game theory to study particle-based vehicle traffic models, contributing to developing a method for measuring the impact of cooperation \emph{versus} non-cooperation on the performance of vehicle traffic systems, when micro-level interactions are taken into account. In order to achieve this aim, the paper is structured as follows: Section \ref{sec:traffic_network_game} introduces the basic model elements of the vehicle traffic game defined in this work, and presents the main features of a resource allocation protocol used to solve route conflicts among the vehicles in a decentralized manner; in Section \ref{sec:game_validation} it is described how the vehicle traffic game was validated according to several aspects, such as overlappings, starvation, and entropy; further, in section \ref{sec:fuel_based_priorities}, a fuel consumption dynamics is introduced in order to represent this important aspect of real vehicles; then, in section \ref{sec:price_of_anarchy}, the main goal of this paper is accomplished, which is to analyze the impact that cooperation and non-cooperation have in the performance of the system; and, finally, section \ref{sec:discussion} discusses the results and how they can be used in further developments.

\section{The traffic network game}  
\label{sec:traffic_network_game}
Let $\mathcal{G = (V,E)}$ be a graph representing a traffic network. In the graph theory jargon, a network node is called a \emph{vertex} and a network link called an \emph{edge}, so these terms are used hereafter. Vehicle agents occupy the vertices in $\mathcal{V}$ and move through the edges in $\mathcal{E}$, in discrete time. Each vehicle $a_i$ starts at a source vertex $s_j$ and has a destination vertex $d_j$. At each time $t_n,n=1,2\dotsc$, a vehicle chooses the next vertex among the neighbors of the current vertex and moves to it between the times $t_n$ and $t_{n+1}$. This assumption of unitary travel time for all edges is used for simplicity, however, this can be generalized to arbitrary times. The choice of a vertex is instantaneous and respects certain restrictions. This assumption is plausible because with autonomous vehicles this choice involves only sensing and computing, so it can be approximately instantaneous.

The graph vertices and edges can be used by any number of vehicles at any time, however, in this game, this is considered an unsafe situation that should be avoided, which we call an \emph{encounter} or, equivalently, \emph{overlapping}. A \emph{vertex encounter} may occur at any time instant. An \textit{edge encounter} is said to happen at edge $e$ at time $t$ when more than one vehicle uses $e$ to move between the times $t_{n-1}$ and $t_n$. Depending on the physical interpretation, an encounter may represent or not a collision, but this definition is not necessary for the present analysis, though it might be needed in future studies. When the vehicle reaches its destination vertex, this vertex is occupied for just one time step, after which the vehicle is removed. These definitions are made for a game where each vehicle is a player having the mission to reach its destination vertex, and encounters are avoided as much as possible, with decisions that can be made either individually, in one scenario, or my means of a central entity, in another scenario.

A rule of this game is that vehicle $a_i$ has a quantity of fuel $\phi_i$, which decreases at a certain rate with each movement, possibly becoming less than enough for making any further move, a situation in which the vehicle has to stop for refueling. This situation is referred to as \textit{starvation}, and is a model feature which is assumed because of the limited planning capabilities which exist in practice: despite the route to the destination is known, it is not possible to fully predict the interference of other traffic onto the own vehicle trajectory (as well as of weather and other external factors, in real systems). 

\subsection{Distributed resource allocation protocol}
\label{sec:resource_allocation_protocol}
The distributed or decentralized resource allocation protocol is essentially a set of rules for each vehicle to find its route in a distributed interaction,  without referring to a central arbitrator or controller, in such a way that it reaches its destination without overlapping with other vehicles along. Another way of thinking of this game is that each vehicle has to solve an online optimization problem \cite{DunkeThesis} for route finding, subject to constraints coming from the resource capacities, from the states of the other vehicles and from the resolution protocol.

A novel protocol of such type was developed for this study, as described in Appendix~\ref{append:protocol_definition}. It is a special type of deferred acceptance matching mechanism \cite{Gale1962, Milgrom2014}, doing matching between vehicles and vertices by means of single bid auctions. Each vehicle uses a priority value which determines its valuations of the resources and, consequently, influences in its bid values. This protocol was validated and applied to some instantiations of the vehicle traffic network game defined above and subjected to performance analysis in several aspects, as described in several parts of this paper. 

\subsection{Example of instantiation of the vehicle game}
\label{sec:game_tetrahedral}
\begin{figure}[h]
	\begin{center}
		\includegraphics[width=0.7\textwidth]{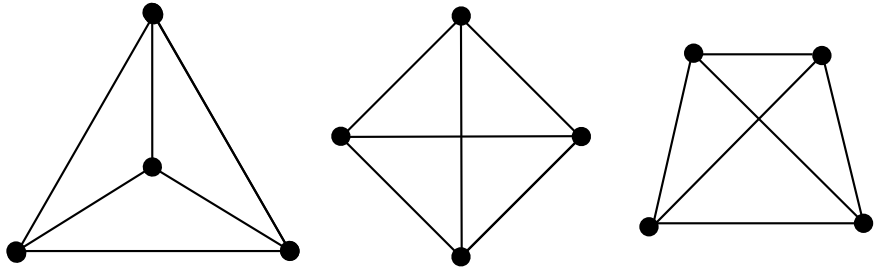}
		\caption{Some projections of the tetrahedral circuit.}
		\label{fig:tetrahedral_graph}
	\end{center}
\end{figure}

Let $\mathcal{G}$ be a tetrahedral graph, which is the maximal complete planar graph and can be depicted as any of the examples in figure~\ref{fig:tetrahedral_graph}, among other forms. In the initial case considered, these edges are bi-directional ($(u,v)=(v,u)$) and have no capacity limitation, however, overlappings must be avoided by the conflict resolution protocol and the vertex choosing strategy. 

\begin{figure}[ht]
	\begin{center}
				\begin{subfigure}{.3\textwidth}
					\includegraphics[]{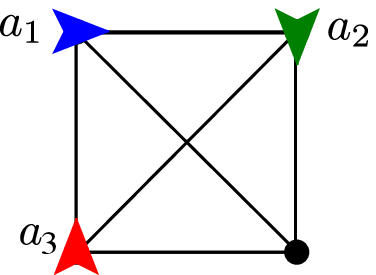}
					\caption{No conflict;}
				\end{subfigure}
					~
				\begin{subfigure}{.3\textwidth}
						\includegraphics[]{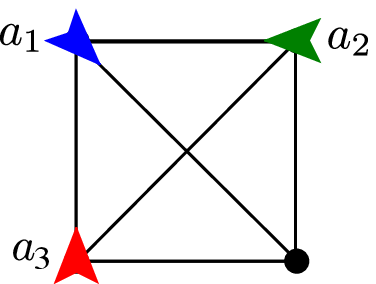}
						\caption{Vertex conflict, $\xi=(a_2,a_3)$;}
				\end{subfigure}
						~
				\begin{subfigure}{.3\textwidth}
						\includegraphics[]{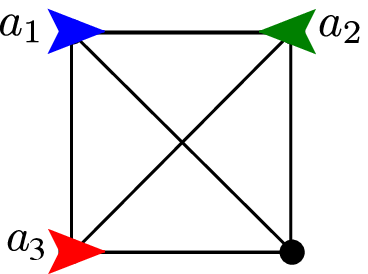}
						\caption{Edge conflict, $\xi=(a_1,a_2)$.}
				\end{subfigure}
							
		\caption{Some initial vehicle configurations in the tetrahedral circuit.}
		\label{fig:tetrahedral_vehicles}
	\end{center}
\end{figure}

Three vehicles are placed at vertices in the beginning of the simulation, without overlappings, and have to reach a destination vertex distinct from its starting position, thus they have a direction, as shown in figure~\ref{fig:tetrahedral_vehicles}. 
		
In the absence of conflicts, it is possible to complete all missions in just one time step. Conflicts, though, may happen, as in figure \ref{fig:tetrahedral_vehicles}.b, in which vehicles $a_2$ and $a_3$ are disputing the same vertex for the next step and, and as in figure \ref{fig:tetrahedral_vehicles}.c, where vehicles  $a_1$ and  $a_2$ are disputing the same edge. In order to solve these conflicts, the resource allocation protocol and the vertex choosing strategy are used, avoiding overlappings in the extent possible and enabling accomplishment of the vehicles´ missions. 

Considering the number of initial game configurations, if the distinction of vehicle ids is disregarded, there are in total 108 distinct cases: 4 possibilities of vehicle positioning at vertices, as there is just one vertex without vehicle each time; in each of these possibilities, each vehicle has 3 possible choices for the next vertex and the destination vertex, which are the same. It is assumed that, in the initial state, the intended next vertex cannot be the current vertex (a vehicle cannot start with the intention of holding its position). As there are three vehicles, the possibilities for all vehicles are multiplied, thereby totaling $3^3$ possibilities for each initial positioning, thus resulting in $4\times3^3=108$ initial configurations. Considering the permutations of 3 distinct vehicle ids, this number is multiplied by $3!$, amounting to $108\times 6=648$ initial configurations. It is assumed that these initial configurations have equal probabilities. 

Each of the initial configurations will open a tree of possible game trajectories. In order to evaluate the effect of the priority relationship established by the vehicles’ priority indices, these trajectory trees were used to explore the entire game state space, with the probabilities of  conflict being calculated at the tree nodes.  An example of a game tree with the possible game trajectories is presented in Appendix~\ref{append:trajectory_tree}.   

\section{Validation of the vehicle traffic game}
\label{sec:game_validation}
Experiments were done to check the various aspects of the game and are reported in Appendix~\ref{append:validation}. Besides the visualization of a trajectory tree example, the following characteristics have been examined:
\begin{itemize}
	\item \textbf{Vehicle overlapping}: despite being hard to guarantee, via an analytical proof, that a distributed resolution protocol such as the one here avoids vehicles overlapping in 100\% of the situations, here it was exhaustively tested and no overlapping was found. 
	\item \textbf{Probability of vehicle starvation} given a limited amount of fuel: this gives some measure of the efficacy of the protocol with various combinations of values of vehicle priorities vector. A vehicle starving before reaching destination generates extra costs and, depending on how the handling of this situation is, may have unsafe consequences.
	\item \textbf{Probability of non-termination} with the hypothesis of unlimited fuel: because the distributed resolution protocol is non-deterministic, it is possible that the game trajectory instantiations become cyclical and, in some cases, the cycles can go on infinitely. Obtaining the probability of non-termination gives a measure of the efficacy of the conflict resolution protocol, although in practice the vehicles will have limited fuel or have to stop for maintenance.
	\item \textbf{Traffic entropy analysis}: measuring the entropy of the traffic configurations generated by different vehicle priority combinations helps to understand the patterns which contribute to increasing or decreasing the performance of the distributed conflict resolution protocol.
	\item \textbf{Hold vs. No-hold}: comparisons were done with some performance indicators between the cases where the vehicles can hold position along time and when they cannot do so. For the present game definitions, the no-hold case was deemed better suited and used for the core performance analyses of this paper.	
\end{itemize}

Based on those findings, the definitions of the game and the resource allocation / conflict resolution protocol were considered consistent, robust and safe for the purposes of this study.

\section{Fuel-based priorities}
\label{sec:fuel_based_priorities}
This section is dedicated to defining and using some game characteristics that express more truly the reality existing in some transportation systems in which fuel consumption plays a prominent role. 
\subsection{A more realistic fuel consumption model} 
\label{subsec:fuel_consumption_model}
Vehicles that carry their own fuel have worse performance when they are loaded with more fuel. This phenomenon is hardly noted by us in our everyday use of ground  transportation, however,  is highly significant for aircraft pilots, race vehicles,  rockets, and others. As the fuel gets consumed along the trip, the vehicle mass diminishes and its performance improves; this way, for a given speed, the fuel consumption rate decreases as the trip progresses. This behavior is represented by the following differential equation:
\begin{equation}
	\label{eq:fuel_diff}
	\dot{\phi}(t)=\rho-\lambda\phi(t)\mathrm{,}
\end{equation}
where $\phi$ is the fuel quantity of the vehicle, $\rho$ is a constant value corresponding to the fuel rate when $\phi\rightarrow 0 $, and $\lambda$ is another constant. This means that the rate of fuel consumption is proportional to the sum of a constant fuel rate, corresponding to the physics of the empty vehicle, and a term proportional to the current amount of fuel. In more informal words, the amount of fuel burnt per second is higher when there is more fuel in the tank to be carried. Underlying this statement, it happens that the power generated by the engines, which is proportional to the fuel burning rate in a given environment condition, is gradually reduced along the trip.

Being a linear ODE, a solution for equation~\ref{eq:fuel_diff} can be easily obtained (e.g., by Laplace transform), thereby allowing to write an explicit formula for the amount of fuel remaining at time $t$:
\begin{equation}
	\label{eq:fuel_resolved}
	\phi(t)=\left(\phi^{\langle 0\rangle}-\frac{\rho}{\lambda}\right)e^{-\lambda t}+\frac{\rho}{\lambda}\mathrm{,}
\end{equation}
where $\phi^{\langle 0\rangle}$ is the initial amount of fuel. When this is given, the formula allows calculating the amount of fuel remaining at each moment in time, until it vanishes. On the other hand, over the assumption of constant speed and constant environmental conditions for the engine, $t$ can be fixed to a value $t_{\min}$ which is the duration of the shortest path between origin and destination, and from equation~\ref{eq:fuel_resolved} it is possible to calculate the minimum amount $\phi^{\langle 0\rangle*}$ of fuel necessary to accomplish the vehicle's mission, supposing no rerouting. 

This fuel consumption model has a direct relationship with the Breguet Range Equation \cite{Ruijgrok2009}, widely used for aviation flight planning, according to the following equations:
\begin{equation}
\label{eq:breguet_lambda}
\lambda=g\frac{\mathrm{SFC}}{(C_L/C_D)}
\end{equation}

\begin{equation}
\label{eq:breguet_rho}
\rho=-m_{\mathrm{min}}g\frac{\mathrm{SFC}}{(C_L/C_D)}
\end{equation}
where $g$ is the gravitational constant and the following parameters determine the aircraft performance characteristics: $m_{\mathrm{min}}$ is the minimum ``flyable'' aircraft mass (the aircraft itself plus the minimum legal amount of fuel and the payload), $C_L$ is the lift coefficient, $C_D$ is the drag coefficient, and SFC is the Specific Fuel Consumption, a measure of the engine efficiency.   

\subsection{Fuel-based priority rule}
\label{subsec:fuel_based_priority}
Unlike section~\ref{sec:game_tetrahedral}, where the priority of each vehicle is constant throughout the game, here fuel and priority are coupled by equations expressing a relationship between each other. For vehicle $i$ at time $t_n$,
\begin{equation}
\label{eq:fuel_based_priority}
w_i^{\langle n\rangle}=\left[\frac{\max\left(0,\phi_i^{\langle n\rangle}-\phi_i^{\langle n\rangle*}\right)}{\tau_i-\phi_i^{\langle n\rangle*}}\right]^\epsilon\mathrm{,}
\end{equation} 
where $\phi_i^{\langle n\rangle*}$ is the exact amount of fuel necessary to reach the destination vertex using the shortest route, and $\tau_i$ is the tank capacity of the vehicle. So, the max numerator in equation~\ref{eq:fuel_based_priority} gives the amount of ``spare'' fuel, and the denominator gives a proxy measure for fuel spent and, consequently, for the past length of the travel. The exponent $\epsilon \geq 0$ provides a non-linear bending towards 0 ($\epsilon > 1$), stressing the importance of numerator, or towards 1 ($\epsilon < 1$), stressing the importance of the denominator. In this study, $\epsilon = 1$ is used throughout.

It is intuitively reasonable to make the vehicle priority proportional to the spare fuel, such that, if the vehicle is low on fuel, it will not make detours; in many types of vehicles, insufficient fuel is a safety issue (e.g. heavier-than-air aerial vehicles). In the denominator, it could be used just the tank capacity as a way to normalize the priority, however, this would have ruled out an important principle of job scheduling, as a traffic system can be seen as a job processing system where each job is a vehicle's mission. If a system has several ongoing tasks, it is most often effective to prioritize jobs with smaller time to completion, because once they finish, resources are freed and the longer jobs can be expedited. This also decreases systemic entropy. As  $\phi_i^{\langle n\rangle*}$ is directly proportional to the distance to destination, subtracting it in the denominator achieves the desired priority effect. 

Other factors that could be taken into account in the priority rule are the payload quantity, payload type, mission type, etc., but they are not considered in the present study.

\subsection{Fuel dispatch strategies and penalty policies}
With the above principles understood, one realizes that the fuel-based priority rule brings no incentive to carry extra fuel. Actually, vehicles carrying extra fuel will not only burn more fuel per unit of time, but also will have higher priority index and, consequently, be more likely to make detours to give way for emptier vehicles. This is very plausible at peer-to-peer conflicts, however, it would be a myopic rationale in the systemic context. If two conflicting vehicles have the minimum fuel, one of them will have to deviate anyway, thus it is guaranteed it will not reach its destination. Based on this reasoning, it is not optimal to depart with the minimum amount of fuel.

For simplicity, there is only one degree of freedom in a vehicle's fuel dispatch strategy: the initial priority index $w_i^{\langle 0\rangle}$, from which is determined the initial amount of fuel $\phi_i^{\langle 0\rangle}$. This determination is based on equation~\ref{eq:fuel_adjusted_to_weight} below, 

\begin{equation}
\label{eq:fuel_adjusted_to_weight}
\phi_i^{\langle 0\rangle} = \min \left[\tau_i,\phi_i^{\langle 0\rangle*}+w_i^{\langle 0\rangle}\left(\tau_i-\phi_i^{\langle 0\rangle*}\right)\right]
\end{equation}
which, in turn, is originated in equation~\ref{eq:fuel_based_priority}. 

In order to avoid introducing more degrees of freedom to the problem, it is assumed that $w_i^{\langle 0\rangle}$ has to be chosen indifferently to the mission routes. Nevertheless, the numerous uncertainties in real operations dilute the importance of prior route knowledge. 

In the model developed here, the penalty policy for landing on an alternate location  (a circuit vertex other than that of the mission destination) has also a simple definition, being the sum of a fixed and a variable component. The fixed one is a fuel-equivalent operational charge  $\widetilde{\Phi}$, and the variable one is $\phi_i^{\langle n\rangle*}$, the minimum amount of fuel needed to reach the original mission destination. In the real world, the costs for a transportation vehicle stopping and re-dispatching at an unintended location are numerous. Firstly, there is energy dissipation in decelerating and energy consumption for accelerating and, in the case of aerial vehicles, there is also descending and climbing. Secondly, refueling (or recharging, in the case of an electrical vehicle) and doing another departure check takes time. Thirdly, the incurred extra time generates extra financial costs from the market discount rate applied to the capital invested in the equipment and from contractual fines. Besides, the delay is likely to cause customer dissatisfaction and many other subjective costs. 

These various cost factors are wrapped up, in this model, in a fuel-equivalent amount, and this allows working with a single and intuitive cost quantity (a negative payoff) for the game outcome. Thus, the fuel penalty $\widetilde{\phi}_i^{\langle n\rangle}$ for the $i$-th vehicle starving at time $t_n$ is defined as 
\begin{equation}
\label{eq:fuel_penalty}
\widetilde{\phi}_i^{\langle n\rangle}=\widetilde{\Phi}+\phi_i^{\langle n\rangle*}
\end{equation}

\section{Analyzing the performance of distribution and cooperation}
\label{sec:price_of_anarchy}
It is sought here to compare the social performance of the game when the utility maximization occurs individually (anarchically or non-cooperatively) and when it occurs cooperatively. The effect of cooperation may be supplanted by a central coordinator, and this case will be analyzed here. If agents are rational, selfish and know how others play, there may be anarchical equilibrium points where each agent's expected individual utility is maximized with respect to the choices available to itself and to the others. There are several variants for the definition of equilibrium, being the Nash's the most used one \cite{Myerson1997}. A Nash equilibrium is the combination of individual strategies from which no agent, having knowledge of other agents' available strategies, can take benefit from unilaterally changing its own strategy. Furthermore, equilibrium theory is a central element of Mathematical Economics and is widely used for analyzing competitive markets such as those of airlines, as done in \cite{Ciliberto2009}, from a more purely economic point of view, and in \cite{Waslander2008}, which considers the competition for traffic resources. The latter work, however, is elaborated on a Eulerian traffic flow model, distinctly from the present work, which uses a Lagrangian approach. Considering the game under study here, it is not possible to state the existence of Nash equilibria, because the strategy parameter $w_i^{\langle 0\rangle}$ is continuous and the payoff function is not. Having the game a negative payoff represented by the total fuel-equivalent cost, this payoff jumps discontinuously between airports/vertices. However, a simplifying assumption is taken, which is to admit only a set of finite values for the strategy parameter, and this allows an initial analysis of game equilibria. 

The vehicle traffic game used to analyze how different strategies perform against the cost logic is very similar to that validated in section \ref{sec:game_tetrahedral}, with the same tetrahedral circuit, three vehicles, resource allocation protocol $\boldsymbol{P}$ and routing strategy. Additionally, the fuel consumption model of section~\ref{subsec:fuel_consumption_model} was used, with $\rho=1$, $\lambda=0.02$, and $\epsilon = 1$, as well as the fuel-based priority assignment of section~\ref{subsec:fuel_based_priority}. The vehicle tank capacity $T$ is defined as 5.0 units of fuel, and the fuel-equivalent penalty $\widetilde{\Phi}$ for starvation is defined as 2.0 units of fuel. Both the individual and collective performance of these two settings were measured and analyzed, as described in the next subsections. 

\subsection{Optimum strategy vector}
\label{subsec:optimum_tetrahedral}
Before defining the discrete values to be used by the agents as the strategy choices, it is worth finding which the optimal social outcome for the game is. Letting $w_i^{\langle 0\rangle}$ vary between 0 and 1 for the three vehicles and using the general purpose optimization algorithm DIRECT \cite{Jones1993} as implemented in \cite{NLOpt}, it was possible to find that the optimum $\boldsymbol{w}^{\langle 0\rangle}$ is $(0,0.51,0.51)$, yielding a collective cost of 4.99 units of fuel. Under the hypothesis of symmetry, the other two permutations of this vector, $(0.51,0,0.51)$ and $(0.51,0.51,0)$, are equally optimal.

\subsection{Discrete-valued strategies}
It can be supposed that an agent (a vehicle) has to choose from a finite set of values of $w_i^{\langle 0\rangle}$ and, for simplicity, the only two scalar values of the optimum vector are used: 0 and 0.51. With this assumption, it is possible to build a payoff matrix of fuel costs incurred, as shown in table~\ref{tab:payoff_matrix_tetra_game}.

\begin{table}[htbp]
\caption{Payoff table of the vehicle traffic game with tetrahedral circuit.}
\label{tab:payoff_matrix_tetra_game}
\centering
\begin{tabularx}{7cm}{|c|c|cc|cc|}
\cline{2-6}
\multicolumn{1}{c|}{}& $w_1^{\langle 0\rangle}=$ &\multicolumn{2}{c|}{0.0}&\multicolumn{2}{c|}{0.51}\\[1.5ex]
\cline{2-6}
\multicolumn{1}{c|}{}& $w_2^{\langle 0\rangle}=$  & 0.0 & 0.51 & 0.0 & 0.51\\[1.5ex]
\cline{1-6}
\multirow{9}{*}{\begin{sideways}$w_3^{\langle 0\rangle}=$\end{sideways}}
& \multirow{4}{*}{\begin{sideways}0.0\end{sideways}} 
   & 2.68 & 2.21 & 2.22 & 1.83\\
&  & 2.60 & 2.22 & 2.21 & 1.83\\
&  & 2.56 & 2.21 & 2.21 & 1.32\\
&  & $\overline{7.84}$ & $\overline{6.64}$ & $\overline{6.64}$ & $\overline{4.99}$ \\
\cline{2-6}
& \multirow{4}{*}{\begin{sideways}0.51\end{sideways}} 
   & 2.21 & 1.32 & 1.83 & 1.87\\
&  & 2.21 & 1.83 & 1.32 & 1.82\\
&  & 2.22 & 1.83 & 1.83 & 1.80\\
&  & $\overline{6.64}$ & $\overline{4.99}$ & $\overline{4.99}$ & $\overline{5.49}$\\
\cline{1-6}
\end{tabularx}
\end{table}

In the top and left cells, this table shows the combination of values of the priority vector $\boldsymbol{w}^{\langle 0\rangle}=\left(w_1^{\langle 0\rangle},w_2^{\langle 0\rangle},w_3^{\langle 0\rangle}\right)$. In the central cells, the table shows the expected costs incurred by each of the vehicles and their total. The order of these cost numbers in these cells is, from top to bottom, $a_1$, $a_2$, $a_3$ (each of individual players' cost) and their total, marked with a horizontal bar on top. It is possible to observe that the lowest possible individual and collective cost occurs at $\boldsymbol{w}^{\langle 0\rangle}=(0,0.51,0.51)$ and its permutations, and the highest collective cost is achieved at $\boldsymbol{w}^{\langle 0\rangle}=(0,0,0)$. Despite the game being conceptually symmetrical, there are small differences among the individual payoffs at the points with equal priorities. The reason for this is that, in case there are two or more simultaneous conflicts to be solved, with vehicles having equal priorities, the implementation of the conflict resolution protocol (in Appendix~\ref{append:protocol_definition}) will use the vehicle id (1,2,3) to select the first conflict to be solved, defining the highest as the most prioritary. It is not possible to randomize this choice, otherwise, the protocol would fail. But these differences are not relevant to the performance analyses which follow.

\subsection{Finding equilibrium and the price of anarchy}
\label{subsec:price_of_anarchy_tetrahedral}

A question of high importance here is if there is some equilibrium point in this game. In order to analyze this, it is important to have in mind that the choice of priority value is simultaneous, with agents having to make their decisions not knowing the opponents' decisions. In this case, a Nash equilibrium exists if there is a tuple of strategies in which each strategy is the best response to the other ones. If the game has only two players, $(A,B)$ is an equilibrium if $A$ is the best response to $B$ and vice versa. If the game has three players, as it is the example developed here, $(A,B,C)$ is an equilibrium if $A$ is the best response of player 1 to player 2 choosing $B$ and player 3 choosing $C$; $B$ is the best response of player 2 to player 1 choosing $A$ and player 3 choosing $C$; and $C$ is the best response of player 3 to player 1 choosing $A$ and player 2 choosing $B$. This is clearer when using \emph{pure} strategies, i.e., when $A$, $B$ and $C$ are deterministic, so that one agent can completely anticipate what the others will play, but this situation cannot be found in the present case.

Reasoning from the viewpoint of an individual agent, without coordination, it is impossible to obtain one of the optimum points $\boldsymbol{w}^{\langle 0\rangle}\in\{(0,0.51,0.51),(0.51,0,0.51),(0.51,0.51,0)\}$ in a deterministic way, because one of them would have to be \emph{chosen} to have $w_i^{\langle 0\rangle}=0$ and the others to 0.51. On the other hand, it is possible to maximize the expected payoff by drawing $w_i^{\langle 0\rangle}$ randomly from a binomial distribution with $\Pr\{0\}=p$ and $\Pr\{0.51\}=1-p$. Doing this and optimizing $p$ either for the individual or the collective payoff, the optimal result is achieved with $p = 4/33 \approxeq 0.12$. With this mixed strategy, the expected individual cost is $1.79$ and the expected collective cost is $5.38$. Being this strategy played by one agent the best response to the same strategy played by another agent, there is no stimulus for an individual agent to deviate from it, so it constitutes an equilibrium point, and it can be demonstrated as the \emph{only} equilibrium point of this game. With these elements, it is possible to evaluate the performance of cooperation. In order to do so, this concept is based on the concept of \emph{price of anarchy} used in Economics.

The price of anarchy can be considered the most popular measure of the inefficiency of equilibria. Formally, the price of anarchy of a game is defined as the ratio between the worst utility value of an equilibrium of the game and that of an optimal outcome \cite{Roughgarden2007}. According to this, we have, at the only equilibrium point, a collective cost of $5.38$ and, at the optimum point, a cost of $4.99$ units of fuel, hence a ratio of $1.0782$ or an excess of 7.82\% over the optimal value. It cannot be considered a huge overhead, but in a low-margin and environmentally critical industry such as transportation businesses, this percentage is very significant. 

In order to close this gap, some coordination among the agents and some altruistic mechanism is needed. The altruistic mechanism could be either enforced or induced, and can succeed only if the game is iterated. If the players have the belief that others can act altruistically, then he or she has the motivation to act altruistically and motivate the others to do so, by seeking a coordination mechanism and complying with it, at the present or future times, decreasing his/her own cost in the long term. Describing it more formally, if the traffic game is repeated infinitely many times and, at some point, an agent may switch between selfish and altruistically, the coordinated choice of $\boldsymbol{w}^{\langle 0\rangle}\in\{(0,0.51,0.51),(0.51,0,0.51),(0.51,0.51,0)\}$ may become a subgame perfect Nash equilibrium. In order to make it possible, an effective coordination protocol has to be established and made feasible in practice. However, there still can be a further optimization if the players change more radically their behavior, as explained below.

\subsection{Comparison with centralized conflict resolution}
\label{subsec:centralized_tetra}
Departing from the game theoretic perspective, the traffic problem here presented can be solved by centralized optimization algorithms, which can achieve global optima and, concerning the collective costs, can beat solutions achieved incrementally by agents with partial information. In the centralized solution developed here, the problem elements differing from the game above are the conflict resolution and the vertex choosing strategy, which become centralized in a single algorithm. In this case, a central traffic control has information about all current vehicle states and their missions and diverts the path of each vehicle in order to avoid encounters and achieve the lowest collective cost. The fuel uplift is still left to the individual vehicles, but the vehicle priorities are dropped because this would restrict the set of solutions available to the central optimizer. This is equivalent to have equal priorities among the vehicles. 

Over these assumptions, brute force optimization is used, i.e., for each combination of initial game configuration and fuel uplifts, all game trajectories generated are visited, and that with the lowest collective cost is chosen in each case. Here, still the fuel uplift has to be done without knowing the initial game configuration, but just the own mission, which always has the same distance. 

Running the DIRECT optimization algorithm in this scenario, the optimal result was achieved when each aircraft loads 3.1 units of fuel. Because there are so many initial game configurations, the optimal collective cost is averaged for them, as in the previous section. Doing this, the average optimal collective cost incurred is 4.07 units of fuel. Because the generation of initial configuration is symmetrical for all vehicles, the expected cost per vehicle is one-third of this value, 1.36. These payoffs are 24\% better than the equilibrium values of the previous section.

Even among centralized solutions, there may be different optimal values depending on how the problem is modeled, which information is used and which degrees of freedom are chosen. Here, there could be a further optimization if, when doing the fuel uplift, before departure, the vehicles used the information of all vehicles' missions. The centralized conflict resolution could be computed in advance and the vehicles would load only the minimum fuel necessary, allowing the vehicles to move at their best efficiency. This scenario was implemented and run for optimization, however, no improvement was obtained, the reasons for this can be conjectured as: i) the cost function is non-linear, non-convex and discontinuous, with continuous input variables, thus the search can only be heuristic and non-exhaustive on the values of the input variables; and ii) the optimality gap is so small that the optimization algorithm used cannot find an improvement in practical time.

\subsection{Larger game: 3$\times$3 grid}
\label{subsec:grid09_game}
\begin{figure}[ht]
	\begin{center}
		\includegraphics[]{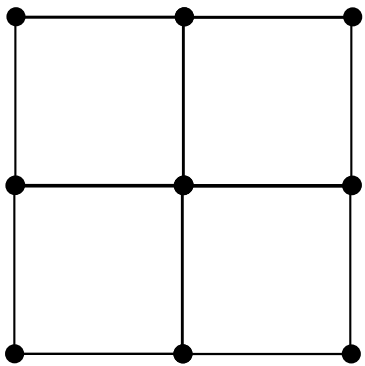}
		\caption{Traffic circuit grid with $3\times 3$ vertices.}
		\label{fig:grid3x3}
	\end{center}
\end{figure}

In order to assess if the results obtained for the tetrahedral circuit are analogous to more complex circuits, the modeling elements used so far are now applied in a larger circuit. In this game instance, the traffic network is a square grid with $3\times 3$ vertices without diagonal edges, as in figure~\ref{fig:grid3x3}. Three vehicles are placed in the network circuit, each at some vertex, without overlapping, and it has to reach some other vertex in the grid. The initial direction is a consequence of using the routing strategy without conflict verification. The resource allocation protocol and the routing strategy remain the same as in the original game from section \ref{sec:game_tetrahedral}. The number of initial configurations of vehicle positioning and directions explodes in relation to the tetrahedral case, reaching 258,048, with the respective trajectory trees considerably larger, because of the larger circuit. For this reason, the vehicle tank capacity was enlarged to $T=10.0$. All the remaining elements of the game are similar to those present in the game previously analyzed.

Initially, it was sought to find the optimum strategy vector $\boldsymbol{w}^{\langle 0\rangle}$, which was found as being $(0,0.32,0.54)$ and permutations, by using the same optimization methods as described in sub-section \ref{subsec:optimum_tetrahedral}. Here, though, each game tree evaluation takes considerable more time, so the optimization algorithm was used in successive stages with human intervention to close up the search intervals. A total of 130 game tree evaluations was considered enough to admit convergence to the vector valuation just referred, which yields a collective cost of 7.78 units of fuel. And, as previously, the combinations obtained with the scalar values of the optimum vector are used to generate a payoff matrix, shown in table \ref{tab:payoff_matrix_grid9_game}.   

\begin{table}[htbp]
	\caption{Payoff table of the vehicle traffic game with a 3$\times$3 grid circuit.}
	\label{tab:payoff_matrix_grid9_game}
	\centering
	\begin{tabularx}{13cm}{|c|c|ccc|ccc|ccc|}
		\cline{2-11}
		\multicolumn{1}{c|}{}& $w_1^{\langle 0\rangle}=$ &\multicolumn{3}{c|}{0.0}&\multicolumn{3}{c|}{0.32}&\multicolumn{3}{c|}{0.54}\\[1.5ex]
		\cline{2-11}
		\multicolumn{1}{c|}{}& $w_2^{\langle 0\rangle}=$  & 0.0 & 0.32 & 0.54 & 0.0 & 0.32 & 0.54 & 0.0 & 0.32 & 0.54\\[1.5ex]
		\cline{1-11}
		\multirow{12}{*}{\begin{sideways}$w_3^{\langle 0\rangle}=$\end{sideways}}
		& \multirow{4}{*}{\begin{sideways}0.0\end{sideways}} 
		&    3.00 & 2.63 & 2.62 & 2.97 & 2.84 & 2.75 & 3.00 & 2.92 & 2.86\\
		&  & 2.98 & 2.97 & 3.00 & 2.63 & 2.84 & 2.92 & 2.62 & 2.75 & 2.86 \\
		&  & 2.97 & 2.63 & 2.62 & 2.63 & 2.13 & 2.11 & 2.62 & 2.11 & 2.10\\
		&  & $\overline{8.94}$ & $\overline{8.23}$ & $\overline{8.23}$ & $\overline{8.23}$ & $\overline{7.80}$ & $\overline{7.78}$ & $\overline{8.23}$ & $\overline{7.78}$ & $\overline{7.82}$\\
		\cline{2-11}
		& \multirow{4}{*}{\begin{sideways}0.32\end{sideways}} 
		&    2.63 & 2.13 & 2.11 & 2.84 & 2.71 & 2.63 & 2.92 & 2.84 & 2.78\\
		&  & 2.63 & 2.84 & 2.92 & 2.13 & 2.70 & 2.84 & 2.11 & 2.63 & 2.78\\
		&  & 2.97 & 2.84 & 2.75 & 2.84 & 2.70 & 2.62 & 2.75 & 2.62 & 2.50\\
		&  & $\overline{8.23}$ & $\overline{7.80}$ & $\overline{7.78}$ & $\overline{7.80}$ & $\overline{8.11}$ & $\overline{8.10}$ & $\overline{7.78}$ & $\overline{8.10}$ & $\overline{8.06}$\\
		\cline{2-11}
		& \multirow{4}{*}{\begin{sideways}0.54\end{sideways}}     
		&    2.62 & 2.11 & 2.10 & 2.75 & 2.63 & 2.50 & 2.86 & 2.78 & 2.74\\
		&  & 2.62 & 2.75 & 2.86 & 2.11 & 2.62 & 2.78 & 2.10 & 2.78 & 2.73\\
		&  & 3.00 & 2.92 & 2.86 & 2.92 & 2.84 & 2.78 & 2.86 & 2.50 & 2.73\\
		&  & $\overline{8.23}$ & $\overline{7.78}$ & $\overline{7.82}$ & $\overline{7.78}$ & $\overline{8.10}$ & $\overline{8.06}$ & $\overline{7.82}$ & $\overline{8.06}$ & $\overline{8.20}$\\
		\cline{1-11}
	\end{tabularx}
\end{table}

As in the tetrahedral version of the game, there is no pure equilibrium, so a mixed strategy is needed. For an agent, $p$ is defined as the probability of drawing 0, $q$ as the probability of drawing 0.32, and $(1-p-q)$ the probability of drawing 0.54; so, it is possible to optimize the expected cost from table \ref{tab:payoff_matrix_grid9_game}, and find that $p=0.22$ and $q=0.45$ provide the minimum cost, $2.67$ for an individual, and $8.00$ for the collective cost. Comparing the latter with the optimum collective cost of $7.78$, one reaches the conclusion that the price of anarchy in this version of the game is $\approx 2.82\%$. This gap is considerably smaller than that of the tetrahedral circuit game and is probably due to the larger number of vertices in the circuit, causing fewer conflicts.  

The centralized version of this scenario was probed too, following the same modifications used in Section~\ref{subsec:centralized_tetra}, with an individual vehicle using information only about its own mission. The expected collective and individual costs in this scenario are 7.26 and 2.42 units of fuel, respectively, an improvement of 9.08\% relatively to the equilibrium result.   

\subsection{Competition analysis}
As long as the traffic game here studied is symmetrical for all vehicles, the individual payoffs at the equilibrium are equal. However, the game may be symmetrical or not depending on the information that each vehicle has available and on the topology of the circuit. Under the symmetry hypothesis, the agents have to make their decisions based only on the information of their own mission and on the knowledge of the circuit. For the tetrahedral circuit, this condition associated with the topology and with the uniform length of the mission paths made that each vehicle saw exactly the same situation at the beginning. In the $3\times 3$ grid version of the game, there could be differentiation based on the geometry and topology of mission path of the own vehicle, although the possibility of using this information, and not only the path length, was not explored, so the game continued symmetrical.

However, if the vehicles have full information of all vehicles' missions before deciding, this makes that each initial game configuration brings a different situation for each vehicle, as if each one were a different game, in which each vehicle may be in advantage, disadvantage or indifferent. Thus the game becomes asymmetrical and it would be very hard to reach agreements \emph{during} the game, if there is no previous agreement or mandatory rule required for entering and staying the game. In real-world transportation systems with multiple agents, asymmetries are common and are partly absorbed by agents and partly mitigated by traffic authorities. Agents can compete for staying at the good side of the asymmetry, but authorities exist to promote a balance among capacity, efficiency, and fairness which maximizes the public good. The formation of coalitions or monopolies would be good for canceling the asymmetries out by cross-subsidization, however, monopolies have inefficiencies of their own type, so they may or may not be a good solution. Thus, when analyzing the performance of traffic conflict resolution among autonomous vehicles, it is important to look at some measure of the fairness of the solutions.

We can say that this game is fair if the incurred costs for the vehicles are close to one another and, in this case, we will measure the equality or inequality of those costs. A popular measure of inequality is the Gini coefficient \cite{Gini}, which is usually applied to monetary income or wealthy of individuals in a society, but can be applied here in a reverse manner to measure differences in the extra fuel cost spent by each vehicle due to conflict resolution in a particular game instance. The Gini coefficient varies between 0, where all individuals have the same income or cost, and 1, for a society with infinite inequality, which is only theoretical because it would need an infinite number of individuals. A practical formula used for calculating is given in~\cite{Sen}, which is used here, having as input the following individual measurement: when a vehicle finishes the game, its fuel cost, added to the starvation penalty, if it happened, is registered; this is the \emph{incurred} fuel cost. The fuel cost which would be the minimum necessary to accomplish the mission without deviations is called the \emph{mission minimum} fuel cost. The difference between the incurred and the mission minimum is called the \emph{excess cost}, and the ratio of the excess cost over the mission minimum is denominated the \emph{excess ratio} and this is the input measure to calculate the Gini coefficient. This measure was chosen because it normalizes the size of the excess cost by the length of the original mission. But the formula of~\cite{Sen} still has a singularity if all individuals have measurement value zero, what may happen in the traffic game. For this exceptional case, it is defined here that the Gini coefficient is zero, meaning total equality. 

The Gini coefficient thus gives a measurement of how much asymmetry there is in the solution of a particular game instance and, consequently, how difficult would be the acceptability of that solution or the underlying resolution rules by individual agents, if the asymmetry persists after consecutive instantiations of the game. The following section summarizes the results of the different scenarios, including the Gini coefficients of each one.

\subsection{Comparative summary of the conflict resolution scenarios}
In order to make clearer the distinction between the non-cooperative and the cooperative scenarios of the vehicle game, described in the previous sections, here follows a summarized explanation. In the non-cooperative scenario, the vehicles choose their priority values regardless of the values which other agents choose. This does not happen in the cooperative cases. In the first cooperative case, the agents have to team up and assign priority values to them according to one of the optimal combinations of the priority vector. This means that they establish a differentiation among them, implying that some agent will achieve lower utility and some will achieve higher utility. Protocols for this coordinated decision making can be found in \cite{VilaplanaThesis,Moniz2009}. This way, the equilibrium would happen only along a sequence of game instantiations, possibly infinite, resulting in what we may call a subgame perfect Nash equilibrium, and a trust building mechanism would be necessary to maintain this equilibrium \cite{Ramchurn2004,Ramchurn2004a}.

In a second cooperative scenario, the traffic system itself is changed, and not only the priority assignment has to be coordinated, but also the definition of the individual vehicle trajectories, which is performed by a single central controller. Still, the central controller can be an external entity, can be one of the vehicles, or can be replicated in every vehicle, but this is not an essential aspect. The essential aspect is that the central controller uses information from the whole scenario and has power over every vehicle, hence this case is said simply \emph{centralized}. There is a very large variety of centralized optimization methods for traffic, for example \cite{Bertsimas1998,Murca201596,CastroFortes2015}; although, in the present paper, brute force search was used because it was enough for the case studies elaborated. As discussed in the introductory section of this paper, centralized strategies are theoretically more efficient than distributed strategies, however, are less robust to perturbations. Anyway, the performance measures of the different types of scenario can be compared based on the results obtained in this study. They are summarized in Table~\ref{tab:performance_summary}.

\begin{table}[htbp]
	\caption{Performance comparison among the different conflict resolution scenarios, with the two different circuits.}
	\label{tab:performance_summary}
	\centering
	\begin{tabular}{C{2.0cm}C{5.5cm}C{3.0cm}C{3.0cm}}
		\hline
		Circuit & Conflict resolution scenario & Collective cost & Gini coefficient \\ \hline \hline
		 & Non-cooperative, distributed (ND) &  5.38 & 0.495 \\
		Tetrahedral & Cooperative, distributed (CD) & 4.99 & 0.409 \\
		 & Cooperative, centralized (CC)& 4.07 & 0.502 \\ \hline
		 & Non-cooperative, distributed (ND) & 8.00 & 0.494 \\
		$3\times 3$ grid & Cooperative, distributed (CD) & 7.78 & 0.416 \\ 
		 & Cooperative, centralized (CC) & 7.26 & 0.580 \\ 
		\hline
	\end{tabular}
\end{table}

In this table, the collective cost values are just the results obtained in the previous sections. For the evaluation of the Gini coefficient, each game instantiation (the game trajectory resulting from an initial configuration) was re-visited and provided individual costs which were then used as inputs. These evaluations are averaged, with weights respected, in order to calculate the measure in the last column.   

The collective cost values are in accordance with the intuition that, the more centralized the resolution is, the better is the collective result and, in a symmetrical society, this would be the best case for the individual. On the other hand, when asymmetry is considered, the Gini coefficient is less intuitive. It is the highest at the centralized scenarios (CC), lowest at the cooperative distributed (CD), and intermediate at the non-cooperative (NC). In an elementary Pareto analysis with the dimensions of cost and equity, we can say that CC and CD are at the efficient frontier, while NC is not, thus only the cooperative scenarios are Pareto-efficient.

\section{Conclusions}
\label{sec:discussion}
A distributed resource allocation protocol was developed in this study with the purpose of studying the performance of a group of vehicles where each one has to travel from point A to B in a traffic circuit, avoiding conflicts in a cooperative, but distributed manner. In order to increase the realism of this performance analysis, elements were introduced to simulate fuel consumption and other costs  associated to the vehicle operation. Assuming that the vehicle priority is determined by the amount of fuel loaded at departure (equation \ref{eq:fuel_based_priority}), both the individual and collective performances of the vehicle traffic game were analyzed according to the choices which the agents can make when loading the fuel. It was considered the non-cooperative case, where the agents do not coordinate themselves to determine the amount of fuel, and the cooperative case, where they coordinate to make this choice. Besides, a derivation of the cooperative case was elaborated to substitute the distributed resource allocation protocol for a centralized algorithm, to serve as an approximation of the best performance theoretically achievable for the traffic system. 

Each of these resolution scenarios was applied in two different circuits, with distinct sizes and topologies. Taking the non-cooperative distributed scenario as the baseline, the costs achieved by the cooperative distributed scenario showed savings of 7.25\% in the tetrahedral circuit and of 2.75\% in the $3\times 3$ grid, and by the cooperative centralized scenario, savings of 24.3\% and 9.08\% in the respective circuits. 

In practice, the cooperative achievement of the optimal cost requires additional communication and either a central control entity or a coordination protocol among the agents. Besides, and perhaps more importantly, this requires policies and means to compensate or mitigate asymmetries. If the asymmetries can be compensated along iterated game matches,  credit/debit systems can be used and trust building mechanisms can help \cite{Ramchurn2004,Ramchurn2004a}. In this scenario, if one agent behaves altruistically in a match, it will stimulate the other agents to do so in the next matches. However, in the real life, the vehicles tend to repeat routes and times, so the conflict configurations naturally repeat and the asymmetries have to be absorbed either by individual adaption or by cross-subsidization. Therefore, one criterion for choosing a conflict resolution strategy is how much asymmetry it can absorb or create when it is used. This study demonstrated that, for distributed strategies, a cooperative one results in less asymmetry than a non-cooperative one, and the centralized strategy results in the highest asymmetry. This deficiency of the centralized strategy can be mitigated if the equity coefficient itself becomes part of the optimization, but this possibility was not explored here.

The model on which this analysis was build is still distant from real vehicle traffic systems, although the analogies sound plausible. In order to increase the realism, the main development to be done is to refine the concept of \emph{resources} and their \emph{capacities}. For example, a loading/offloading station can be represented by a graph node with capacities, e.g., a maximum number of simultaneous vehicles, a number of vehicles processed per unit of time, etc.; the same for a spatial traffic sector and route segments. In some system concept definitions \cite{Asadi}, the real resources can be managed by means of virtualization, such as that the access to a spatial sector is granted by allocating a time slot in a communication channel operated by the STDMA protocol \cite{Amouris}. Thus the disputed resources become time slots, separating the resource allocation problem into two layers: one, where the resource allocation protocol from here can be used to allocate time slots in a communication channel, and another, where Model-Predictive Control (MPC) \cite{Garcia1989} algorithms are used to provide safe trajectories within a traffic sector. The advantage of using the algorithm of Appendix~\ref{append:protocol_definition} in combination with the approach of \cite{Asadi} is to improve optimization across spatial sectors (one sector becomes a node in the conceptual circuit used in the present paper), because the aircraft priority ordering would be more stable through sector transitions, hence decreasing system entropy and rendering the benefits associated with this decrease.

Besides all this, there is the strong trend of vehicles becoming electric, both the ground and the aerial ones. The equations used in Section~\ref{sec:fuel_based_priorities} for modeling the vehicle energy consumption cannot be used for electric vehicles, however, the influence of energy management seems equally or more important. Electric vehicles do not have significant mass variation along the trip, but the battery output voltage decays, thus the performance worsens along the trip. This effect is opposite to that of fuel-based vehicles and, for best performance, the electric vehicle should have the maximum charge at each departure. In this case, the charge time becomes critical and, beyond this, the battery charge cycle history impacts the useful life of the battery, which is costly. Still, replacing batteries or changing battery size are design alternatives, but there are also costs to it. Studying the influence of these factors on traffic performance sounds a very interesting topic.

It was previously demonstrated that the use of distinct vehicle priorities improves the performance of MPC algorithms \cite{Chaloulos,Siva} as well as more elementary conflict resolution algorithms \cite{RomanideOliveira2012}, but such studies did not explore the foundations of the effects of priorities, as done in this paper. Here, this effect has been formulated as the opposite of the price of anarchy, as understood in game theory applied to Economics, thus enabling a more systematic method to evaluate its impact on the system performance. Beyond this, it was attempted here to analyze the impact of using the knowledge of each other vehicles' routes on the traffic performance, however, the optimization algorithm and the computational resources were not good enough to allow some conclusion on it. So, this remains as another topic for future research. The performance analysis of distributed conflict resolution with the fundaments of this paper could also be applied to hierarchical combinations of centralized and distributed control, as it would be a case where the traffic is divided into cells or sectors and, within each sector, the vehicles would refer to a central controller, while there could be decentralization among the sector controllers to control their boundary flows.   

From the practical point of view, the results of \cite{Chaloulos,Siva,RomanideOliveira2012} show that better communication and collaboration mechanisms improve the performance of a traffic system, and this paper presents a systematic way to explain and measure this improvement.  However, the assignment of priorities is just one way of comparing non-cooperative or anarchical interactions, at one side, to cooperative or centralized interactions, on the other side. These two opposite concepts can be compared without the use of priorities \cite{Dimarogonas2003, Bakule2008, Dimarogonas2010}, although the priority variation mechanism facilitates a continuous scale between non-cooperativeness and full cooperativeness.

\section*{References}
\bibliographystyle{elsarticle-num}
\bibliography{./game_paper2015_2}

\pagebreak
\appendix
\renewcommand{\thesection}{\Alph{section}}
\counterwithin{figure}{section}
\section*{Appendices}

\section{List of mathematical symbols}

\begin{table}[htp]
	\caption {Main symbols used in the basic definition of the vehicle traffic game.} \label{tab:symbol_definitions} 
	\begin{center}
		\begin{tabular}{l p{10.5cm}}
			\hline
			Symbolic definition & Description \\ \hline
			$N_a^{\langle n\rangle}$ & Number of vehicles at time $t_n$;  \\ 
			$\mathcal{A}_n=\{a_i:i=1,2,...,N_a^{\langle n\rangle}\}$ & Set of vehicles at time $t_n$;  \\
			$w_i\in [0,1]$ & (min-) Priority of the vehicle $a_i$; \\
			$u_i\in\mathcal{V}$ & Current vertex of vehicle $a_i$; \\
			$\nu_i\in\mathcal{V}$ & Next vertex chosen by vehicle $a_i$; \\
			$d_i\in\mathcal{V}$& Destination vertex of vehicle $a_i$; \\
			$\phi_i^{\langle n\rangle} \in\mathbb{R}^+$ & Fuel amount of vehicle $a_i$ at time $t_n$;\\
			$s_i^{\langle n\rangle}=(w_i,u_i,\nu_i,d_i,\phi_i)^{\langle n\rangle}\in \boldsymbol{S}_n$ & State of vehicle $a_i$ at time $t_n$; \\ 
			$\boldsymbol{S}_n=\left\{s_i^{\langle n\rangle}:i=1,...,\left|\mathcal{A}_n\right|\right\}$ & State of the game at time $t_n$;\\
			$\mathcal{C}_n=\left\{c_k^{\langle n\rangle}:k=1,...,N_c^{\langle n\rangle}\right\}$ & Set of conflicts at time $t_n$; \\
			$c_k^{\langle n\rangle}=\left(r_k,\xi_k\right)^{\langle n\rangle}\in \mathcal{C}_n$ & The $k$-th conflict at time $t_n$; \\
			$\boldsymbol{r}_k\in \mathcal{G}$ & Set of resources (a vertex, an edge or both) being disputed in the conflict $c_k^{\langle n\rangle}$; \\
			$\xi_k\subseteq \mathcal{A}_n$ & Set of vehicles disputing $\boldsymbol{r}_k$ in $c_k^{\langle n\rangle}$; \\
			$I:\cdot\mapsto 2^{\mathbb{N}}$ & Indices function, retrieves the set of indices of the objects in a set; \\
			$\omega_k=\min\limits_{i\in I\left(\xi_k\right)}(w_i)$ & Priority of the conflict $c_k^{\langle n\rangle}$; \\ 
			$\mathcal{R}_n\subseteq \mathcal{G}$ & Set of allocated resources at time $t_n$;\\ 
			$L_n:\mathcal{R}_n\mapsto 2^{\mathcal{A}_n}$ & Resource allocation at time $t_n$, to be realized at $t_{n+1}$; \\
			$\tau_i$ & Vehicle tank capacity. \\
			\hline
		\end{tabular}
	\end{center}
\end{table}

\section{Definition of the distributed conflict resolution protocol}
\label{append:protocol_definition}
It is assumed that the resolution protocol is the same for all vehicles, however, each vehicle has a different priority $w_i$. Another assumption is that all vehicles symmetrically know the next vertices $\nu_i, i=1,...,\left|\mathcal{A}_n\right|$, being intended by itself and by the other vehicles. The definition of this allocation protocol is given in the frame ``Algorithm~\ref{alg:conflict_resolution_P}''.
The functions \textsc{\ref{alg:conflict_aware_vertex_choice}} and \textsc{\ref{alg:bid_response_delay}}, used in the algorithm definition, will be explained along the next subsections. The most important to have in mind is that this protocol is meant to be executed and entirely terminated within the time window corresponding to the discrete step of time $t_n$, that is: assuming that a finite number of protocol iterations happen, the protocol will never last until $t_{n+1}$. This is a reasonable hypothesis when applied to physical vehicles, because usually, the computing and communication time is negligible when compared to the movement of physical bodies. 

It is also important observing that the protocol never obliges a vehicle to choose some vertex or edge, allowing free will in this choice. However, the vector $\boldsymbol{w}=\{w_i\}_{i=1,...,N_a^{\langle n\rangle}}$ establishes a priority for each vehicle and, consequently, to their associated conflicts. These priorities help to define the order in which the vehicles make their choices and register them in the allocation map $L_n$. This allocation protocol is guaranteed to terminate in a finite number of steps because each vehicle has a hard deadline to choose its next route segment, however, it is not guaranteed to be free of encounters.  

Another key aspect of this protocol it that has an auction for deciding which vehicle will be the first to alternate its path in case of conflict. It is a single bid, worst price auction, where the ``winner'' vehicle is actually the loser, because, assuming that the bids are issued at approximately the same time, the smallest $\Delta_i$ will be the first to alternate path, according to that logic. This feature of the protocol makes it a special type of deferred acceptance matching mechanism \cite{Gale1962, Milgrom2014}, with matching between vehicles and vertices. 

In order to analyze the behavior of this resource allocation protocol, it is worthwhile having a clear and direct relation between the vehicles' priorities and the probability of losing the alternation auction, so as that the vehicle with the highest priority value $w_i$ (the least prioritary in the order) has the highest probability of alternating. Thus, taking the normalized vector of priorities $\boldsymbol{W}$ obtained in line~\ref{alg:protocol_auction-line} of algorithm~\ref{alg:conflict_resolution_P}, each of its components $ W_i$ is defined to be the probability of vehicle $a_i$ losing the alternation auction. The relationship that has to be imposed between $ W_i$ and the stochastic distribution of $\Delta_i$ is analyzed in the following subsection.

\begin{algorithm}[!htp]
	\caption{Resource allocation protocol $\boldsymbol{P}$ for a vehicle $a_m$ at time $t_n$}
	\label{alg:conflict_resolution_P}
	\begin{algorithmic}[1]
		\State continue $\leftarrow$ \textbf{true};
		\While {continue \textbf{and} $L_n[\mathcal{R}_n]\neq\mathcal{A}_n$}
		\State Choose $\nu_m$ based on the own routing strategy and broadcast it;
		\State Determine $\mathcal{C}_n$ based on $\nu_i,i\in I(\mathcal{A}_n)$;
		\If{$\bigwedge\limits_{k\in I\left(\mathcal{C}_n\right),m\in\{1,...,N_a^{\langle n\rangle}\}}\left(\left(a_m\notin \xi_k\right)\wedge
			\left(w_m<\omega_k)\right)\right)$} 
		\State \Call{Allocate}{$m,\nu_m,\mathcal{R}_n,L_n$};
		\State continue $\leftarrow$ \textbf{false}; 
		\Else
		\State $j\leftarrow \argmin_{k\in I\left(\mathcal{C}_n\right)}(\omega_k)$;
		\If {$a_m\in \xi_j$}
		\State $\boldsymbol{W}\leftarrow$ \Call{NormalizedPriorities}{$I(\xi_j)$} ;
		\label{alg:protocol_auction-line}
		\State $\Delta_m\leftarrow$ \Call{BidResponseDelay}{$\boldsymbol{W},m$};
		\State $\tau_m\leftarrow t_n+\Delta_m$;
		\State \textbf{wait} until other vehicle has changed vertex choice \textbf{or} $t\geq\tau_m$;
		\If{$t\geq\tau_m$}
		\State $\nu_m\leftarrow$ \Call{ConflictAwareVertexChoice}{$\boldsymbol{r}_j,\mathcal{R}_n,m$};
		\State			\Call{Allocate}{$m,\nu_m,\mathcal{R}_n,L_n$};
		\State Broadcast the new values of $(\nu_m,\mathcal{R}_n,L_n)$;
		\State continue $\leftarrow$ \textbf{false};
		\EndIf
		\EndIf 
		\EndIf
		\EndWhile
		\Procedure{Allocate}{$m,v,\mathcal{R}_n,L_n$}
		\State $\mathcal{R}_n \leftarrow \mathcal{R}_n\cup\left\{v,\left(u_m,v\right)\right\}$;
		\State $L_n(v)\leftarrow L_n\left((u_m,v)\right)\leftarrow \left\{a_m\right\}$;
		\EndProcedure
		\Function{NormalizedPriorities}{$\mathcal{K}$}
		\State Create the vector $\boldsymbol{W}$ s.t.  $ W_i= w_{k_i},\>k_i\in \mathcal{K},\>i=1,...,|\mathcal{K}|$;
		\State $\alpha\leftarrow \left(\sum_{i\in\mathcal{K}} W_i\right)^{-1}$;
		\State Create the vector $\overline{\boldsymbol{W}}$ s.t., initially,  $\overline {W}_i= |\mathcal{K}|^{-1},\>i=1,...,|\mathcal{K}|$;
		\If {$\boldsymbol{W>0}$}
		\For {$i\leftarrow 1$ to $|\mathcal{K}|$}
		\State $\overline {W}_i\leftarrow \alpha W_i$;
		\EndFor
		\EndIf
		\Return $\overline{\boldsymbol{W}}$;
		\EndFunction
	\end{algorithmic}
\end{algorithm}

\subsection{Path alternation auction}\label{sec:path_choosing_auction}
The goal here is to simulate a sealed first-price auction game, where each of the $n$ players has winning probability $ W_i$ (obviously, $\sum_{i=1}^{n} W_i=1$), and their bid functions $b_i:[0,1] \mapsto\mathbb{R}^+$ have to be calculated to meet the $ W_i$; $b_i$ is a non-decreasing function of the player's valuation $v_i$. Supposing that the value $v_i\in[0,1]$ of each player is stochastically distributed according to a probability density function $g_i$, how can the bid functions $b_i(v_i)$ be determined to match the $ W_i$? 

Without loss of generality, it can be assumed that $b_1(1)\leq b_2(1)\leq ... \leq  b_{N_a^{\langle n\rangle}}(1)$. For a better understanding of the problem, initially it is assumed that $N_a^{\langle n\rangle}=2$. Being $ W_2$ the probability of the player 2 winning, this event can be partitioned in two sub-events: 
\begin{equation}
W_2=\Pr\{b_1(v_1)<b_2(v_2)\}
\end{equation}
\begin{equation}
\label{eq:p_2}
=\Pr\{b_1(v_1)<b_2(v_2)\wedge b_2(v_2)<b_1(1)\}+\Pr\{b_1(v_1)<b_2(v_2)\wedge b_2(v_2)\geq b_1(1)\}
\end{equation}
and each of the terms can be treated separately. The second one is simpler, because the $b_i$ functions are non-decreasing and, as consequence, one event in the conjunction is implied by  the other (approximately, because there is a non-strict inequality; however, it can be assumed that probabilities of exact values are zero). Hence, the implied event can be eliminated:
\begin{equation}
\label{eq:prob_b2_higher_than_all}
\Pr\{b_1(v_1)<b_2(v_2)\wedge b_2(v_2)\geq b_1(1)\}=\Pr\{b_2(v_2)\geq b_1(1)\}
\end{equation}
These probabilities can be calculated using the density functions $g_i$, as long as the events to be measured determine the limits of integration. For (\ref{eq:prob_b2_higher_than_all}), the calculation is:
\begin{equation}
\Pr\{b_2(v_2)\geq b_1(1)\}=\int_{b_2^{-1}(b_1(1))}^{1}\!g_2(v_2)\mathrm{d}v_2
\end{equation}
The first term of (\ref{eq:p_2}) is a little more elaborate and needs a double integral on $v_1$ and $v_2$:
\begin{equation}
\label{eq:prob_b2_lower_than_sup_b1}
\Pr\{b_1(v_1)<b_2(v_2)\wedge b_2(v_2)<b_1(1)\}=\int_{0}^{b_2^{-1}(b_1(1))}\int_{0}^{b_1^{-1}(b_2(v_2))}g_2(v_2)g_1(v_1)\mathrm{d}v_1\mathrm{d}v_2
\end{equation}
Thus the equation for $ W_2$ described by the pdf's is:
\begin{equation}
\label{eq:p_2_integrals}
W_2=\int_{0}^{b_2^{-1}(b_1(1))}\int_{0}^{b_1^{-1}(b_2(v_2))}g_2(v_2)g_1(v_1)\mathrm{d}v_1\mathrm{d}v_2\!+\!\int_{b_2^{-1}(b_1(1))}^{1}\!g_2(v_2)\mathrm{d}v_2\mathrm{.}
\end{equation}
Complementarily, the obtention of $ W_1$ initiates by defining
\begin{equation}
W_1=\Pr\{b_1(v_1)>b_2(v_2)\wedge b_2(v_2)<b_1(1)\}\mathrm{,}
\end{equation}
because $\Pr\{b_1(v_1)>b_2(v_2)\wedge b_2(v_2)\geq b_1(1)\}=0$. From that, one obtains
\begin{equation}
W_1=\int_{0}^1\int_{0}^{b_2^{-1}(b_1(v_1))}g_1(v_1)g_2(v_2)\mathrm{d}v_2\mathrm{d}v_1
\end{equation}
This deduction can be generalized for any number $m$ in a set $\{1,2,...,N_a^{\langle n\rangle}\}$, and the outline of this generalization is: start by defining
\begin{equation}
\label{eq:p_n_notation}
P_{m|q}=
\Pr\left\{\left(\bigwedge_{r\neq m}b_m(v_m)> b_r(v_r)\right)\bigwedge b_{q-1}(1)\leq b_m(v_m) < b_q(1)\right\} 
\end{equation}
where, by definition, $b_0(1)=0$, and the event being measured implies that $b_r(v_r)<b_q(1),\forall r$. The rightmost event in (\ref{eq:p_n_notation}), when varied in $q$, defines a partition of the probability space, so it is possible to define the probability $ W_i$ of a player as
\begin{equation}
\label{eq:p_m_summation}
W_m=\sum\limits_{q=1}^{m} P_{m|q}\mathrm{,}
\end{equation}
having in mind that $\Pr\{b_m(v_m)\geq b_q(1)\}=0$ for $q>m$. This fact has to be used when defining the limits of the nested integrals of the $g_i$'s.

In order to better grasp how this formulation can be concretized, the pdf's can be defined as $g_i=1,\quad\forall i$, corresponding to uniform distributions in the interval $[0,1]$, and the bid functions defined as $b_i(v_i)=B_i v_i$, with $B_i$ scalar constant in the same interval $[0,1]$. This conducts to  perhaps the simplest way to answer to the question on how to define the bid functions $b_i$ when the winning probabilities $ W_i$ are given. However, the equations elaborated above are explicit for $ W_i$ given $b_i$, so if the formulation is inverted, i.e., find $b_i$ given $ W_i$, as it was in the question initially posed, the equations have to be gathered in a system of equations that, when solved, allow to determine $B_i$'s for the given $ W_i$'s. Besides that, one of the $B_i$'s has to be arbitrated, because, despite the fact that there are $N_a^{\langle n\rangle}$ instances of equation~\ref{eq:p_m_summation}, the actual number of independent instances is $N_a^{\langle n\rangle}-1$ since, as $\boldsymbol{W}$ is a vector of probabilities, every instance can be substituted by $ W_i=1-\sum_{j\neq i} W_j$, making it dependent on all others. 

The mathematical formulation developed in this section enables defining the function \textsc{\ref{alg:bid_response_delay}} that is necessary in the definition of the resource allocation protocol $\boldsymbol{P}$. The definition shown in the frame ``Algorithm~2'' is rather generic but its implementation is deducible from the explanations in this section. This function is needed only theoretically in order to demonstrate that the protocol $\boldsymbol{P}$ is implementable as an actually decentralized protocol. However, the simulations performed for this paper are serial and need only the vector $\boldsymbol{W}$. As these simulations explore the whole game trajectory space in a probabilistic way, and the exact values of $\beta_m$ are not strictly necessary for determining the vehicles' trajectories, the results do not lose fidelity.

\begin{algorithm}[!htp]
	\caption{Bid function for the alternation response delay}
	\begin{algorithmic}[1]
		\Function{BidResponseDelay}{$\boldsymbol{W},m$} 	\funclabel{alg:bid_response_delay}
		\State Let $\beta_m$ follow an uniform distribution with limits $(0,B_m]$;
		\State Determine $B_m$ from the system of equations based on  equation~\ref{eq:p_m_summation} and rationale in sub-section \ref{sec:path_choosing_auction};
		\State Draw $\beta_m\sim U(0,B_m]$;
		\State \Return $1-\beta_m$;
		\EndFunction
	\end{algorithmic}
\end{algorithm}

\subsection{Routing strategy}
A very simple routing strategy could be: for the next vertex, choose the one which is the first in the shortest path to the destination and, in case of any conflict, all involved vehicles must start a roundabout maneuver and then finding an ``exit'' to the destination. This definition works also as a resource allocation protocol but, despite apparently effective, there could be overlapping of roundabouts and this situation would require higher level coordination. Although a multi-level coordination algorithm such as \cite{Rinaldi2015}, the present study aims at not requiring hierarchical decision making and, under specific assumptions \cite{Blom2014}, purely distributed strategies can be more effective in solving conflicts. As the present study focuses on basic principles, the elaboration of complex cognitive strategies for the agents is avoided. A rather simple routing strategy is defined here, where the look-ahead time for conflict detection is just one time step. Because of this one-step approach, it is denominated a \emph{vertex choosing strategy}, although each vehicle $a_m$ still has an ulterior goal to accomplish, determined by its destination vertex $d_m$. This strategy is defined by means of the function \textsc{\ref{alg:conflict_aware_vertex_choice}} shown below. 

\begin{algorithm}[!htp]
	\caption{Routing strategy $\mathfrak{N}$}
	\label{alg:routing_strategy}
	\begin{algorithmic}[1]
		\Function{ConflictAwareVertexChoice}{$\boldsymbol{r},\mathcal{R},m$} 	\funclabel{alg:conflict_aware_vertex_choice}
		\State Choose $\nu_m$ as the first vertex in the shortest path between $u_m$ and $d_m$;
		\If {$\nu_m\in \boldsymbol{r}$ \textbf{or} $(u_m,\nu_m)\in \boldsymbol{r}$}
		\State $\nu_m\leftarrow$ \Call{Alternate}{$\mathcal{R},m,\nu_m$};  
		\EndIf
		\State \Return $\nu_m$;
		\EndFunction
		\Function{Alternate}{$\mathcal{R}_n,m,\nu$} 	\funclabel{alg:alternate}
		\State $\mathcal{I}\leftarrow \{(u_m,v)\in\mathcal{E}\}/\left(\mathcal{R}_n\cup\{(u_m,\nu)\}\right)$;
		\If {$\mathcal{I}\neq\emptyset$}
		\State Among the paths which start with an edge in $\mathcal{I}$ and lead to $d_m$, choose the shortest one;
		\State Take $e_{\min}\in\mathcal{I}$ which is the first edge of the shortest path;
		\State \Return the tip vertex of $e_{\min}$ which is not $u_m$;
		\Else
		\State \Return $\nu$;
		\EndIf
		\EndFunction
	\end{algorithmic}
\end{algorithm}

Two or more vehicles can occupy the same vertex simultaneously, however, in the present vehicle game, this is considered an unsafe situation that should be avoided, situation which is called a \emph{vertex encounter}. Distinctly, an \emph{edge encounter} is said to happen at edge $e$ and at time $t_n$ when more than one vehicle uses $e$ to move between the times $t_{n-1}$ and $t_n$. Depending on the interpretation, an encounter may be associated with a collision, but this is a physical concept which is not in the scope of this analysis.

In order to validate the conflict resolution protocol and obtain a better understanding of how the protocol influences the dynamic of the vehicle game, a basic example is presented in the next section.

\section{Experiments for validation of the vehicle traffic game}
\label{append:validation}

\subsection{Example of game trajectory tree}
\label{append:trajectory_tree}
\begin{figure}[ht!]
	\begin{center}
		\includegraphics{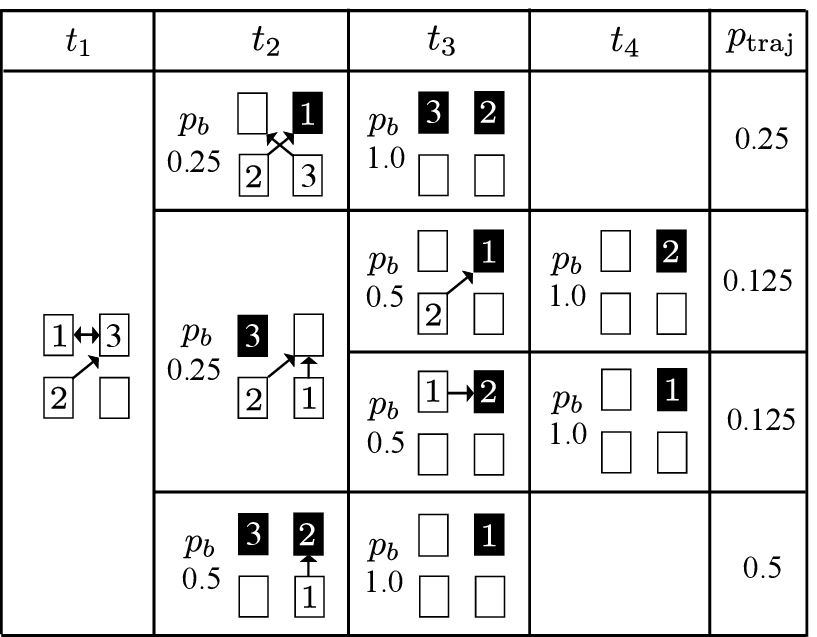}
		\caption{Example of game trajectory tree, with $\boldsymbol{w}=(0.5,0.5,0.5)$ and initial state as depicted in the first column. Time is represented from left to right.}
		\label{fig:trajectory_tree}
	\end{center}
\end{figure}

An example of game trajectory tree for the game with the tetrahedral instantiated in Section~\ref{sec:game_tetrahedral} is given in figure~\ref{fig:trajectory_tree}, with priority vector $\boldsymbol{w}=(0.5,0.5,0.5)$ for the vehicles.  

This tree corresponds to the initial game configuration shown in the first column. For the same value of $\boldsymbol{w}$, there are multiple trees, each one corresponding to an initial configuration. For this example, one of these initial configurations was arbitrarily selected. The columns of the table correspond to successive time steps, and the rows correspond to the trajectories. The initial state is shown on the left, and it may generate one of the three states in the second column, and so continue the game trajectories to the right. Each cell of the table represents a possible state of the vehicle game and is associated with a node of the tree of game trajectories, thus each table cell is referred to simply as a \emph{node}. In each node, the game state uses the following convention: the rectangles represent the circuit vertices, and the number inside the rectangle represents a vehicle, that is, a vehicle id (1, 2, 3). The arrows represent the \emph{intended} route of a vehicle, which in this particular version of the game is just one hop between vertices. Thus, the arrows point to each vehicle's $\nu_i$ (the next vertex, as in the definitions of the algorithm). 

In the initial configuration of this version of the game, at $t_1$, the destination vertex $d_i$ is equivalent to $\nu_i$, therefore, for succinctness, there is no explicit indication of this fact. Thus, at the instant  $t_1$, vehicle 1 is at the top left vertex and intends to go to the top right vertex; similarly, vehicle 2 is at the bottom left vertex and intends to go to the top right vertex; and vehicle 3 is at the top right vertex and intends to go to the top left vertex. Thus, there is an edge conflict between vehicles 1 and 3, because both intend to use the same edge at the same time; and a vertex conflict between vehicles 1 and 2, because both intend to occupy the same vertex on the top right. The resource allocation protocol will be run by the vehicles and they may change their intents to avoid conflicts. The vehicles will move according to the changed intent and, because this allocation protocol is non-deterministic, tree branches are possible from this initial state, shown in the column $t_2$.  In each of the nodes of the tree corresponding to the times $t_2$ through $t_4$, the value of $p_b$ corresponds to the probability of that branch being executed after the parent branch at its left is executed.

If a vehicle reaches its destination vertex, this fact is indicated by the solid black background on the vertex box. For example, in the upper node of the column $t_2$, vehicle 1 reached its destination, thus it is removed when time advances to the next time step. Following the sequence on the upper branch, vehicles 2 and 3 reach their respective destination vertices at time $t_3$ and the game finishes at that branch. The last column of the table shows $p_{\mathrm{traj}}$, which is the probability of that full trajectory being executed, corresponding to the product of the values of $p_b$ throughout the nodes it passes. It is important observing that the values of $p_b$ do not trivially follow the values of $\boldsymbol{w}$, because a game node may have multiple conflicts, as shown in the figure, and they are solved separately by the resource allocation protocol. The internal steps of conflict resolution at each step are not shown in figure~\ref{fig:trajectory_tree}, otherwise, it would become too complicated. 

The analyses presented in the following subsections rely on generating certain values for the priority vector $\boldsymbol{w}$ and, for each of them, generating and exploring the trajectory tree corresponding to every initial game configuration. 

\subsection{Starvation analysis}

In this example the game begins with each vehicle having a fuel quantity $\phi_i^{\langle 0\rangle}=5,\>\forall i$. Each movement to an adjacent vertex consumes 1 unit of fuel, as well as for holding the current vertex, and the game is executed along the probabilistic trajectory trees, as explained above, for the values of priority vectors generated, enabling the estimation of the probability of encounter or starvation as function of the vector of vehicle priorities. In order to avoid that starvations generate encounters due to a stuck vehicle, the starved vehicle is automatically removed once its fuel becomes zero. This is plausible in the case of aerial vehicles, which cannot stay aloft without fuel.

\begin{figure}[ht]
	\begin{center}
		\includegraphics[width=\textwidth]{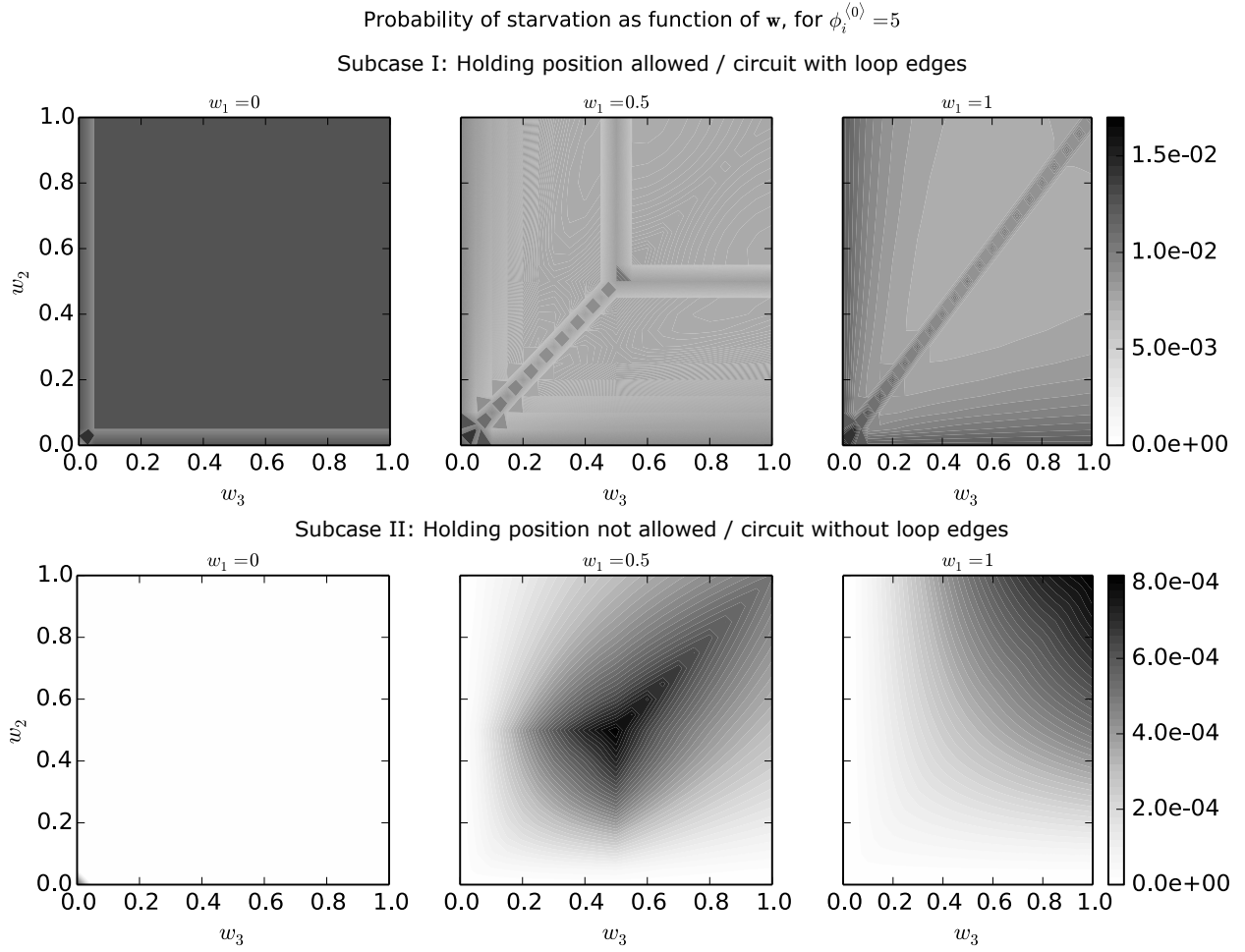}
		\caption{Probability of vehicle starvation, as function of the vector of vehicle probabilities, in the case of limited fuel, initially $\phi_i^{\langle 0\rangle}=5,\>\forall i$.}
		\label{fig:tetrahedral_starvation}
	\end{center}
\end{figure}

The computation of the probabilities of starvation for this scenario is shown in figure~\ref{fig:tetrahedral_starvation}, where the probabilities are measured per vehicle. This computation explores every possible game trajectory in a game tree corresponding to a vector $\boldsymbol{w}$ and, using the $\boldsymbol{w}$ to calculate the probability of each trajectory, keeps initially a separate account for each vehicle id, then averages these accounts for the three vehicles with equal weights, generating a point in the graph corresponding to an instance of $\boldsymbol{w}$. The same evaluation was performed for the Subcase~I, where the circuit has loop edges at the vertices, thus allowing the vehicles to hold positions between successive time steps, ant for the Subcase~II, where the circuit has no loop edges and, thus, the vehicles are not allowed to hold the same position between different time steps. 

One of the clearest facts from observing figure~\ref{fig:tetrahedral_starvation} is that Subcase~II has much lower probability of fuel starvation everywhere. The reasons for this difference will be discussed in the following sections. Another feature noticeable in this figure is that the cases with two or more vehicles having the same priority value $w_i$ are the most critical ones, with higher probabilities of starvation. This feature is similar for both subcases I and II, but with some differences. In Subcase~II, if one or two vehicles have priority zero, there occurs no starvation. Besides, the lines of equality are more pronounced in Subcase~I. Behind this observation is a characteristic of the resource allocation protocol, by which if two or more vehicles have first priority, they become equally important, and this introduces some disorder. However, the case without holdings is a little more robust to that situation, thus the regions of equality are more diffuse.   

\subsection{Overlapping analysis}

In order to do a rigorous evaluation of the effectiveness of the conflict resolution protocol in avoiding vehicle overlappings, the whole space of combinations of the priority vector $\boldsymbol{w}$ must be taken into account. On the one hand, it is impossible to cover every combination, because it has continuous components. One the other hand, though, values in this space can be put in equivalence classes, within each of them any value generates a certain structural configuration of nodes of the trajectory tree or a permutation of nodes with equivalent structure (some tree mirrorings preserve equivalence). Although the vehicle priorities in the nodes vary from one tree to another generated in the same equivalence class, these priorities are not relevant to check whether or not there are  overlappings or starvations in the nodes, what matters is their rank order, and so it is enough to analyze just one vector value inside the equivalence class. 

According to the resource allocation protocol $\boldsymbol{P}$ and the routing strategy  $\mathfrak{N}$ (Algorithm \ref{alg:routing_strategy}), the influence of the vehicles’ priority vector $\boldsymbol{w}$ in defining the trajectory trees is distinguished by  the combination of cardinalities of its component values, according to the classes shown in table~\ref{tab:equivalence_classes}.

\begin{table}[ht]
	\caption {Equivalence classes of vehicle priority vectors regarding the trajectory trees generated.} \label{tab:equivalence_classes} 
	\begin{center}
		\begin{tabular}{cl}
			\hline
			Class no. & Members of the equivalence class of $\boldsymbol{w}$,\\ 
			& with $0<$ $x$ $<$ $y$ $<$ $z$ $\leq$ 1 for each line below.\\ \hline
			1 &\{(0, 0, 0), ($x$, $x$, $z$), ($z$, $z$, $z$)\} \\	
			2*&\{(0, 0, $z$)\}  and permutations.\\
			2 &\{($x$, $z$, $x$), ($z$, $x$, $x$)\}.\\
			3*&\{($0$, $y$, $z$), ($0$, $z$, $z$)\} and permutations. \\
			3 &\{($x$, $y$, $z$),($x$, $z$, $z$)\} and permutations. \\
			\hline
		\end{tabular}
	\end{center}
\end{table}

Because the resource allocation protocol necessitates a tie-breaking rule when two vehicles have the same priority value, there is one case where not only the cardinality of the priority values matters, but also the order in which they appear, and this is the reason why the case  ($x$, $x$, $z$) is in a category distinct than that of  ($x$, $z$, $x$) and  ($z$, $x$, $x$). According to the routing strategy for individual vehicles, it makes difference in some of these combinations if the lowest priority values are 0 or not. In the case of (0, 0, 0) and ($z$, $z$, $z$), this doesn't matter because the normalization makes them all with equal alternation probability; but, in the case of classes 3* and 3, the difference between 0 and $x$ does matter because, in the case of priority value 0, the probability that the corresponding vehicle loses the route alternation auction is zero, therefore it will never alternate vertex; regarding classes 2* and 2, the distinction in the conflict resolution is a little more complicated because, when a conflict occurs between the two vehicles with priority value 0, both vehicles may alternate, but in a conflict involving one vehicle with priority 0 and another with higher priority value, the vehicle with 0 will never alternate. Besides these distinctions, it seems a little odd that ($x$, $x$, $z$) is in equivalence class 1, but this happens because the vehicle ids are used to tie break resolution of simultaneous conflicts with vehicles with same priority values, with lower ids being solved first, making the first two cases in class 1 structurally equivalent to the third. Examples of concrete class members are shown in table \ref{tab:trajectory_lengths}.

Thus, having 5 equivalence classes of game trajectory tree, and multiplying this number by the 648 initial vehicle configurations, the resulting 3,240 trajectory trees generated were exhaustively examined in search of overlappings, and none of such was found. A similar verification was performed without holding position / loop edges and similarly no overlapping was found.

\subsection{Non-termination analysis}

It was not presented a proof that the routing strategy defined above terminates, i.e., that all vehicles are guaranteed to reach destination at the end of the game. This strategy is invariant for any time step and individual player, therefore it is natural to expect that in some cases the game will enter into a perpetual cycle and not terminate. Indeed it was observed that some sequences of route alternations of the vehicles are cyclical. In such cases, it happens that  all vehicles successively alternate path, due to the existence of chained conflicts that, because of the distributed nature of the protocol, cannot be solved with the guarantee that no new conflicts will be generated. In order to evaluate the probability with which this occurs, the trajectory trees are once again employed and, when a game state is repeated, the game trajectory is terminated at this node. Strictly speaking, these “pruned” trees do not allow evaluating the exact probability of infinite trajectories, because at any non-terminal node of the tree, the result of further alternation auctions may lead to terminal nodes, and it is hard to check if this probability is actually computable. On the other hand, admitting that the probability of an infinite trajectory is equal or lower than the probability of reaching the first repeated node in the trajectory tree, it is possible to obtain upper bounds for probabilities of infinite trajectories. However, it is recommendable to know the maximum length of trajectories without cycle in order to assess if it is computationally safe to completely drop the restriction on fuel for obtaining the trajectory trees. For the case without holding / loop edges, there are 648 possible states for three vehicles in the tetrahedral circuit network, which is the number of initial game configurations, previously explained. For the case with holding / loop edges, there are 1536 possible states for three vehicles in the circuit network. This number is higher because the loop edges can be used after the game starts ($4^3$ combinations of vehicle intent instead of $3^3$). 

\begin{figure}[ht]
	\begin{center}
		\includegraphics[width=\textwidth]{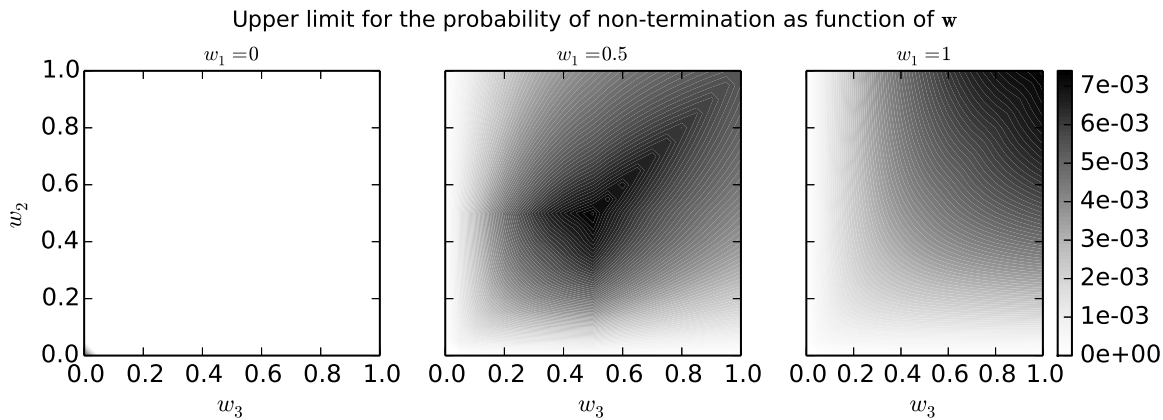}
		\caption{Upper limit for the probability of non-termination of the game as function of the vector of vehicle priorities, in the case of unlimited fuel; holding position not allowed / no loop edges in the circuit.}
		\label{fig:tetrahedral_non_termination}
	\end{center}
\end{figure}

Following the same reasoning, when a vehicle leaves the game after reaching its destination, leaving two vehicles, there may be more 108 states without loop edges and 192 states with loop edges in the circuit. When the next vehicle leaves and just one vehicle remains, the routing strategy does not allow holding position, in which case the latter would go directly to the destination vertex in one step, therefore there is just one possible state. So, the upper bound for the length of a non-cycling trajectory is $648+108+1=757$ for the circuit without loop edges and $1536+192+1 = 1729$ with loop edges. Thus, leaving the amount of fuel of each vehicle equivalent to these amounts, in the respective cases, is equivalent to having unlimited fuel. Doing so for the case without loop edges, the computation of the upper limits of the probabilities of non-termination is obtained as in figure~\ref{fig:tetrahedral_non_termination}.

As it can be observed, the aspect of the shadows is similar to that of figure~\ref{fig:tetrahedral_starvation}, Subcase~II, but with higher values. This allows concluding that trajectory cycles happen with a significant frequency and it is very likely that they are the causes of vehicle starvation observed in the case of limited fuel. The next section helps to shed more light on this relationship. 

It was not possible to perform a similar analysis for the case with holding position / loop edges, because the computing time exploded. Only for a single value of the priority vector $\boldsymbol{w}$, the computation went on for more than 10 hours without generating the probability value for that point, even using the lower limit of 757 units of fuel, so it was interrupted and deemed unfeasible. The reasons behind this problem are related to the discussion of the following section.

\subsection{Finding a measure of effectiveness of the joint routing strategy}
\label{subsec:entropy}
The routing strategy shown above uses a low degree of intelligence, in the sense that it does not explore future unfoldings of the decisions taken at present. If knowledge of future states of the game is to be used \cite{Rinaldi2015}, this would require more computational power, more software development effort and, in many cases, the stochastic nature of the game being played makes impossible to completely eliminate the uncertainties. The analyses of encounter, starvation and non-termination rates, performed in the previous sections, are computationally costly and targeted on specific \emph{effects} of the routing strategy associated with priority vectors, still not elucidating common causes of these effects. As usual in science, finding common causes of distinctly observed effects help not only the comprehension of underlying phenomena, but also to find systematic methods of improving the performance of an engineering system.  

The first step towards finding a general analysis rule is to check the distribution of game trajectory lengths associated with each initial configuration. As pointed out previously, the trajectory trees can be grouped into equivalence classes for ranges of vehicle priority vectors, as shown in table~\ref{tab:equivalence_classes}. One representative for each of these classes was generated, using elements from $\{0,0.5,1\}^3$. Obtaining the simulation trees for these representatives and registering their maximum lengths, it was possible to come up with the data in table~\ref{tab:trajectory_lengths}.  

A possible measure of strategy performance is Shannon's information entropy applied to the game trajectories generated by an initial game configuration, given by the following formula:
\begin{equation}
\label{eq:entropy}
-\sum\limits_i p_i\log_2 p_i\mathrm{,}
\end{equation}
where each $i$ corresponds to a single game trajectory, and $p_i$ the probability of that trajectory being executed. This calculation was performed for the different classes of priority vector $\boldsymbol{w}$ and shown in table~\ref{tab:trajectory_lengths}. Fixing a value assigned to $\boldsymbol{w}$, shown in the second column of table~\ref{tab:trajectory_lengths}, the  calculation of the entropy according to equation~\ref{eq:entropy} was done for each of the 648 initial game configurations, and their maximum value was taken, and this resulted in the values shown in the two cases of the column ``Entropy'' in the same  table~\ref{tab:trajectory_lengths}. The case ``Hold'' corresponds to the case where the game allows holding position (i.e. the circuit has loop edges) and the case ``No hold'' corresponds to the opposite case. All the values of $\boldsymbol{w}$ in the same class produce the same result. Another information present in the table is the maximum length of non-cycling trajectories, in the third column and, besides, if that class of $\boldsymbol{w}$ produces infinite, cycling trajectories, this is indicated with ``Yes'' or ``No'' in the column ``Has cycles?''. These results are also distinguished for the cases ``Hold'' and ``No hold''. 

Analyzing the results in table~\ref{tab:trajectory_lengths}, it is possible to observe two important facts: first, that the ``No hold'' cases have shorter trajectories and lower entropy values; and, second, that the entropy value is correlated with the  trajectory lengths and with the existence of cycles. Lower entropy values have shorter lengths and, in some of the cases ``Hold", the lowest entropies are associated with no cycles in the trajectories.

\begin{table}[ht]
	\caption {Equivalence classes of priorities vectors and their respective trajectory lengths and entropies} \label{tab:trajectory_lengths} 
	\renewcommand{\arraystretch}{1.2}
	\begin{center}
		\begin{tabular}{C{1.2cm}C{6.0cm}C{1.0cm}C{1.0cm}C{1.0cm}C{1.0cm}C{1.0cm}C{1.0cm}}
			\hline Class & \multirow{3}{*}{$\boldsymbol{w}$ class members} & \multicolumn{2}{c}{\small{Max. non-cycling}} & \multicolumn{2}{c}{\multirow{2}{*}{Has cycles?}} & \multicolumn{2}{c}{\multirow{2}{*}{Entropy}} \\
			from & & \multicolumn{2}{c}{\small{trajectory length}} & & & &\\ \cline{3-8}
			Table~\ref{tab:equivalence_classes}& & Hold & No hold & Hold & No hold & Hold & No hold \\ \hline\hline
			1& $\{(0,0,0), (.5,.5,.5), (.5,.5,1), (1,1,1)\}$ & 9 & 7 & Yes & Yes & 3.585 & 2.585\\ \hline	
			2*& $\begin{array}{c}\{(0,0,.5), (0,.5,0), (.5,0,0), (0, 0, 1),\\\> (0, 1, 0), (1, 0, 0)\}\end{array}$ & 8 & 6 & Yes & No & 2.375 & 1.000\\ \hline	
			2 & $\begin{array}{c}\{(.5,1,.5),(1,.5,.5)\}
			\end{array}$ & 9 & 7 & Yes & Yes & 3.391 & 2.459\\ 	\hline
			3* & $\begin{array}{c}\{(0, .5, .5), (0, .5, 1), (0, 1, .5), (0, 1, 1),\\\> (.5, 0, .5), (.5, 0, 1), (1, 0, .5), (1, 0, 1),\\\> (.5, .5, 0), (.5, 1, 0), (1, .5, 0), (1, 1, 0)\}\end{array}$ & 6 & 5 & Yes & No & 2.000 & 1.000\\ \hline	
			3 & $\begin{array}{c}\{(0.5, 1, 1), (1, 0.5, 1), (1, 1, 0.5)\}\end{array}$ & 9 & 7 & Yes & Yes & 3.418 & 2.457\\			\hline
		\end{tabular}
	\end{center}
\end{table}

Because the case ``No hold'' demonstrated better results  in the several analyses presented so far, it was chosen for the next stages of this study, and the case ``Hold'' is disregarded for the core analyses of this paper. Intuitively speaking, the presence of extra edges in the circuit opens more possibilities of maneuvers (i.e. more branches in the game trajectory trees) and, without extra vertices, this brings more disorder to the game, because the game is, by definition, decentralized and non-deterministic. The entropy measurements seem to confirm this intuition. This could be different with more coordinated and smarter resource allocation strategies, however, this would go out of the scope of the present study. 

It is worth remarking that care has to be taken when using the concept of entropy, because this game has irreversible states and thus its entropy should not be understood in the classical sense \cite{Hoyst2009, Lieb2013}. Notwithstanding, the definitions and analysis of this section are considered enough to illustrate the usefulness of the concept of entropy. Some good references on the application of entropy to transportation systems are~\cite{Boyce1980,Christodolou}.

\end{document}